\documentclass[%
 reprint,
 amsmath,amssymb,
 aps,
pra,
]{revtex4-2}

\usepackage{graphicx,xcolor}% Include figure files
\usepackage{dcolumn}% Align table columns on decimal point
\usepackage{bm}% bold math
\usepackage[colorlinks = true,
            linkcolor = blue,
            urlcolor  = blue,
            citecolor = blue,
            anchorcolor = blue]{hyperref}
\usepackage{xcolor}
\definecolor{red}{RGB}{0,0,0}
\begin{document}

\preprint{APS/123-QED}

\title{\Large Topological signatures of collective dynamics\\ and turbulent-like energy cascades in  apolar active granular matter}% Force line breaks with \\
\author{Zihan Zheng$^{1,2}$}
\author{Cunyuan Jiang$^{1,3,4}$}
\author{Yangrui Chen$^{1,2}$}
\author{Matteo Baggioli$^{1,3,4}$}
\email{Corresponding Author: b.matteo@sjtu.edu.cn}
\author{Jie Zhang$^{1,2}$}
\email{Corresponding Author: jiezhang2012@sjtu.edu.cn }
\address{$^1$School of Physics and Astronomy, Shanghai Jiao Tong University, Shanghai 200240, China}
\address{$^2$Institute of Natural Sciences, Shanghai Jiao Tong University, Shanghai 200240, China}
\address{$^3$Wilczek Quantum Center, School of Physics and Astronomy, Shanghai Jiao Tong University, Shanghai 200240, China}
\address{$^4$Shanghai Research Center for Quantum Sciences, Shanghai 201315,China}

\date{\today}% It is always \today, today,

\begin{abstract}
Active matter refers to a broad class of non-equilibrium systems where energy is continuously injected at the level of individual ``particles." These systems exhibit emergent collective behaviors that have no direct thermal-equilibrium counterpart. Their scale ranges from micrometer-sized swarms of bacteria to meter-scale human crowds. In recent years, the role of topology and self-propelled topological defects in active systems has garnered significant attention, particularly in polar and nematic active matter. Building on these ideas, we investigate emergent collective dynamics in apolar active granular fluids. Using isotropic granular vibrators as a model experimental system of apolar active Ornstein-Uhlenbeck particles in a dry environment, we uncover a distinctive three-stage time evolution arising from the intricate interplay between activity and inelastic interactions. By analyzing the statistics, spatial correlations, and dynamics of vortex-like topological defects in the displacement vector field, we demonstrate their ability to describe this intrinsic collective motion. Furthermore, associated to these topological defects, we reveal the onset of a turbulent-like inverse energy cascade, where kinetic energy transfers across different length scales over time. As the system evolves, the power scaling of the energy transfer increases with the duration of observation. Our findings show that topological concepts can be extended to the nonequilibrium dynamics of apolar active matter, revealing a direct link between microscopic topological processes and emergent large-scale behaviors in active granular fluids that lack both a well-defined direction of motion and an intrinsic axis of orientation at the particle scale.
\end{abstract}

\maketitle

%\tableofcontents

\section*{\label{sec:level1}Introduction}
Living and nonliving out-of-equilibrium systems in which the motion of the individual constituents is propelled by an energy input that acts individually and independently on each of them are broadly classified under the common umbrella of \textit{active matter}. Research in this field has expanded so rapidly in recent years that it now requires ``a review of reviews” \cite{vrugt2024review}. Active matter spans an extensive range of scales and appears in a wide class of systems from the bats in \textit{Batman Returns} movie \cite{10.1145/37401.37406} to human crowds partying during traditional celebrations \cite{Gu2025}.

Activity inherently disrupts energy conservation and detailed balance, challenging fundamental principles of thermodynamics and statistical physics \cite{annurev:/content/journals/10.1146/annurev-conmatphys-070909-104101}. Moreover, the interplay between activity and intrinsic interactions in active systems leads to novel non-equilibrium phases of matter, characterized by emergent collective dynamics and spontaneous spatiotemporal patterns \cite{PhysRevX.12.010501}.

In contrast to living systems, granular materials are often considered outside the realm of active substances.
Granular matter consists of numerous grains, typically larger than $100$ $\mu m$, where thermal fluctuations are negligible \cite{RevModPhys.71.S374, RevModPhys-Jaeger, andreotti2013granular}. A pile of sand serves as a prototypical example. Over the past several decades, extensive research has explored the unique properties of granular matter as an athermal amorphous solid, focusing on various fundamental aspects.
Key areas of study include packing properties \cite{Aste-PRE-05, Torquato-MRJ}, granular statistical mechanics \cite{edwards1989theory, Henkse-PRL, RevModPhys-Makse}, jamming transitions and their associated marginal features \cite{Corey-PRE-03, annurev-Liu, van2009jamming, charbonneau2014fractal, charbonneau2017glass, wang2022experimental, PhysRevLett-wyart, annurev:/content/journals/10.1146/annurev-conmatphys-031214-014614, bi2011jamming, pan2023review}, and inhomogeneous internal stress transmission \cite{majmudar2005contact, wang2020connecting, PhysRevLett-Nampoothiri, li2024dynamic}.
Although inherently athermal, granular systems display complex rheological and mechanical properties when subjected to external driving forces, such as shear \cite{midi, jop2006constitutive, Kamrin-PRL, PhysRevLett-Rietz, kou2017granular, shang2024yielding}.

More recently, activity has been introduced into quasi-two-dimensional {\color{red} (2D)} granular systems by placing a layer of granular particles on a vibrating surface, effectively supplying energy to individual particles and allowing them to behave as self-propelling units. A wide range of intriguing phenomena has been observed in these systems, including large-scale collective motion in vibrated polar disks \cite{deseigne2010collective} and tapered rods \cite{Kumar2014}, giant number density fluctuations in \\ swarming granular rods \cite{narayan2007long}, dispersionless vibrational modes \cite{PhysRevLett.133.188302} and gapped shear waves \cite{jiang2024experimentalobservationgappedshear} in bidisperse apolar Brownian vibrators, as well as high-energy velocity tails in granular spinners \cite{PhysRevLett-Thorsten} and monodisperse apolar Brownian vibrators \cite{PhysRevE.106.L052903}. Additionally, flocking behavior has been observed in monodisperse apolar Brownian vibrators \cite{Chen2024}. These systems, commonly referred to as \textit{active granular matter}, form the central focus of our work.

\begin{figure*}[htbp]
   \centering
    \includegraphics[width=0.9\linewidth]{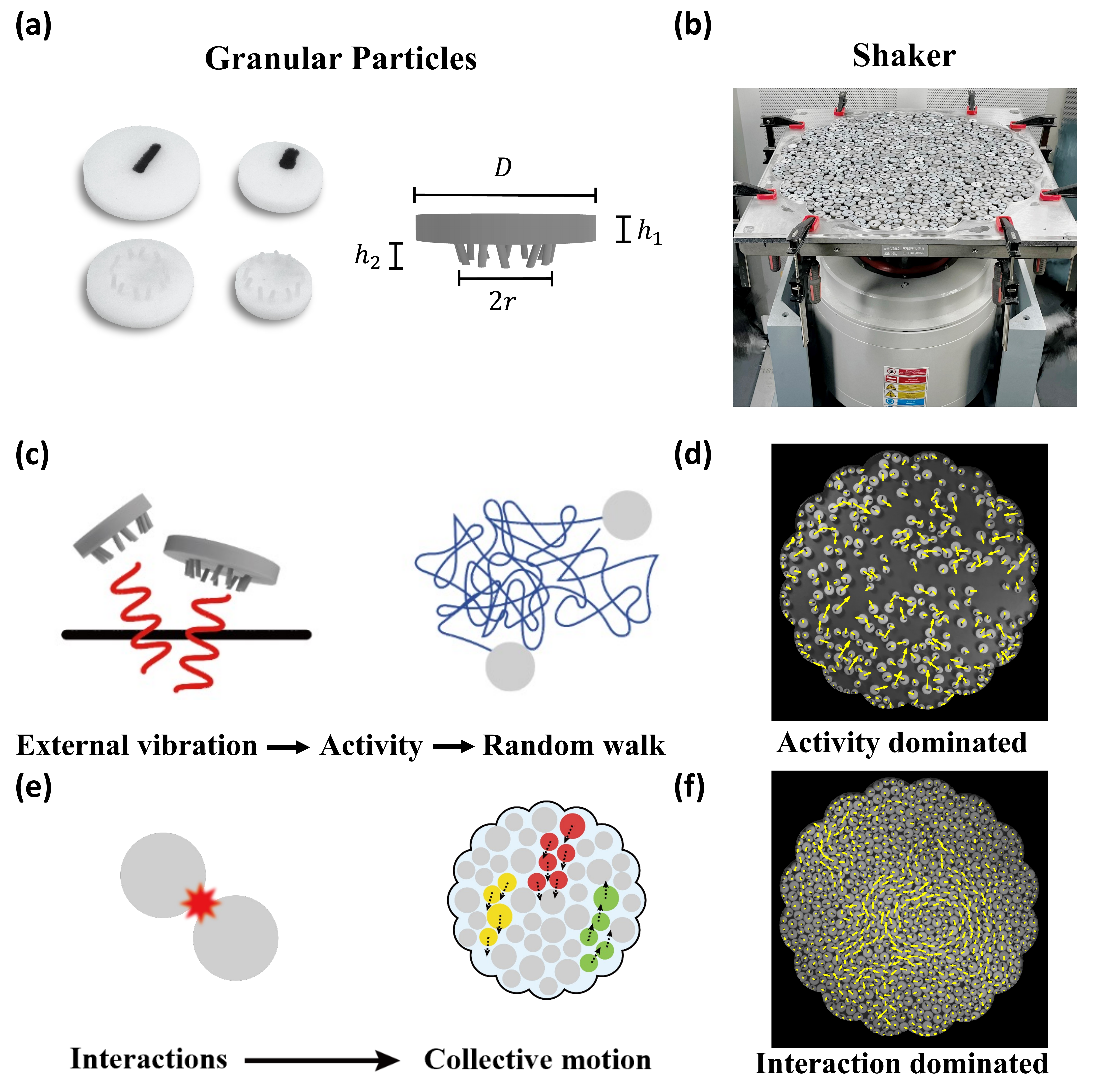}
    \caption{\label{FigE1} \textbf{Experimental setup and underlying physics.} \textbf{(a)} Images of the large and small active \color{red}isotropic granular \color{black}particles used in our experiment. \textit{Left:} Photos showing the top and bottom views of both particle types. \textit{Right:} A schematic diagram illustrating the particle structure with relevant dimensions. \textbf{(b)} An image of the electromagnetic shaker, with a layer of bi-disperse particles placed on an aluminum plate mounted on the shaker. \textbf{(c)} When subjected to vertical vibrations, particles gain kinetic energy individually. This activity induces random collisions, primarily between one of their legs and the bottom plate, leading to isotropic random walk motion in the horizontal 2D plane. \textbf{(d)} Displacement vector field of bi-disperse particles at a low packing fraction ($\phi = 0.309$),  {\color{red} obtained at $t_0=0.025\,s$ and $\tau=500\,s$, with their lengths rescaled by a factor of 0.1 for clarity,} where activity dominates leading to uncorrelated gas-like dynamics. \textbf{(e)} A schematic diagram illustrating how collective dynamics emerge from particle-particle inelastic interactions that become dominant at high packing fractions. \textbf{(f)} Displacement field of bi-disperse particles at a high packing fraction ($\phi = 0.822$),   {\color{red} obtained at $t_0=0.025\,s$ and $\tau=500\,s$, with their lengths rescaled by a factor of 5 for clarity,} exhibiting collective vortex-like structures driven by interparticle interactions. {\color{red}Here, $t_0$ denotes a specific initial reference time, and $\tau$ represents the duration of the observation period $[t_0,t_0+\tau]$.}}
\end{figure*}

Topology has emerged as a powerful tool for characterizing active media, with applications spanning phase transition, material design and biological systems \cite{PhysRevLett-Erwin,Shankar2022,TUBIANA20241,Copenhagen2021,Maroudas-Sacks2021,Meacock2021,doi:10.1126/sciadv.abk2712,Tan2020,Kawaguchi2017}. In particular, topological defects provide a framework for understanding collective dynamics and relaxation mechanisms in active systems \cite{PhysRevX.11.031069,Shi2013,PhysRevX.14.041006}. Much of this research has focused on active polar fluids and nematics \cite{doi:10.1098/rsta.2013.0365}, revealing striking analogies with non-Hermitian condensed matter systems \cite{Ota,sone2024hermitian} and offering new perspectives on active fluid behavior. In such fluids, topological defects act as quasiparticles driven by internal active stresses, generating chaotic flows and turbulence \cite{PhysRevX.5.031003,Shi2013,Sanchez2012,PhysRevLett.110.228101,PhysRevE.88.050502,PhysRevLett.121.108002,PhysRevX.9.041047,doi:10.1126/science.1254784, duclos2017topological}. Beyond active turbulence, defects can be manipulated to perform computations and transmit information, suggesting novel applications in active matter research \cite{doi:10.1073/pnas.2400933121,doi:10.1126/sciadv.abg9060,doi:10.1126/sciadv.abe8494,saw2017topological,guillamat2016control}.

Historically, topology has played a central role in studying polar and nematic active systems, where anisotropic shapes or directional self-propulsion define preferred orientations. Familiar concepts from liquid crystals \cite{Fumeron2023} apply directly in these contexts. Collective motion, typically associated with self-aligning interactions—as exemplified by the Vicsek model \cite{PhysRevLett.75.1226}—was long thought to be restricted to such systems. However, recent theoretical work \cite{PhysRevLett.130.148202}, later confirmed experimentally \cite{Chen2024}, demonstrated that flocking and large-scale motion can also arise in active granular systems without self-aligning interactions. In these systems, effective attractive forces emerge from inelastic collisions, driving collective behavior. \color{red}Still it remains unclear \cite{mahault2023comment}, whether these systems exhibit all the characteristic features of flocking.\color{black}

This work aims to extend the role of topology in apolar \color{red} isotropic \color{black}active granular systems without self-aligning interactions, proving a direct connection between topological defect dynamics, large-scale collective motion, and turbulent-like structures.

\section*{Active granular fluids and single particle dynamics}
In Fig.\ref{FigE1}, we present the most salient features of our experimental setup and the essential ingredients to understand the underlying physics. 

The microscopic constituents in our experimental platform are isotropic granular particles that were 3D printed using a resin material. Each particle consists of a disk-shaped cap with $12$ staggered legs attached underneath, as shown in Fig.~\ref{FigE1}(a). We utilize bidisperse particles with diameter of \( D_l = 22.4 \) mm for large particles and \( D_s = 16 \) mm for small particles.
The other geometric characteristics are the same for large and small particles, namely disk thickness (\( h_1 = 3 \) mm), leg height (\( h_2 = 3 \) mm), and placement of $12$ staggered legs along a concentric circle of radius \( r = 5.6 \) mm. {\color{red}Each leg is inclined inward at an angle of 18.4°, alternating with a deviation of ±38.5° from the particle’s mid-plane. The mixing ratio between large and small particles is set to \( 1:2 \) to suppress crystallization.} Given their size ratio (\( D_l: D_s = 1.4:1 \)), the total area occupied by large and small particles remains equal, maintaining a \( 1:1 \) area ratio in the whole system.

A single layer of particles is then placed on an aluminum plate within a flower-shaped region mounted on top of an electromagnetic shaker, see Fig.\ref{FigE1}(b), {\color{red} designed to suppress collective rotation along the boundary and to facilitate reinjection of particles near the edges back into the system \cite{Dauchot_PRL}. A detailed discussion of the effect of the boundary shape on the final vortex configuration is provided in the Supplementary Information (SI).} The electromagnetic shaker generates a vertical sinusoidal vibration at $100$ Hz, driving the entire system at the individual particle level with a maximum acceleration of \(2.5g\), where \(g\) is the gravitational acceleration. A Basler CCD camera (not shown in the image) is positioned above the experimental platform and continuously records particle configurations for an hour at a rate of $40$ frames per second. Fig.\ref{FigE1}(d, f) provides two images of the particles captured by the camera at different packing fractions.

Before each experiment, the vibration platform was carefully leveled to suppress gravitational drift. Levelness was confirmed by achieving a uniform particle distribution at low packing fractions after two hours of continuous vibration, well beyond the characteristic drift time of a few seconds. The system was then vibrated for two hours prior to data collection to establish a steady initial state, ensuring reproducible and well-controlled conditions.

During an experimental run, each particle individually receives kinetic energy from the shaker's vertical vibration and dissipates it through collisions with other particles and with the plate mounted on the shaker.
Particle collisions are inelastic, characterized by a restitution coefficient of $\epsilon \approx 0.39$ (see SI for measurement details). The static friction coefficient between the particles and the plate is $\mu_s \approx 0.42$ \cite{Chen2024}. The collision between a granular vibrator and the plate acts as the source of the active force. Under external vibration, collisions between each granular vibrator and the plate are effectively collisions between individual legs and the plate. In each event, typically only one leg contacts the substrate. Due to the inclined geometry, every collision imparts a net displacement to the particle in a specific direction, creating a propulsion mechanism that mimics an active force.

Once the energy injection and the dissipation balance on average, a non-equilibrium steady state is reached. 
The system's complex dynamics and collective motions arise from the interplay between individual particle activity and interparticle inelastic interactions, as illustrated in the schematic representations in Fig.\ref{FigE1}(c) and (e).

\begin{figure*}[htbp]
   \centering
    \includegraphics[width=1.0\linewidth]{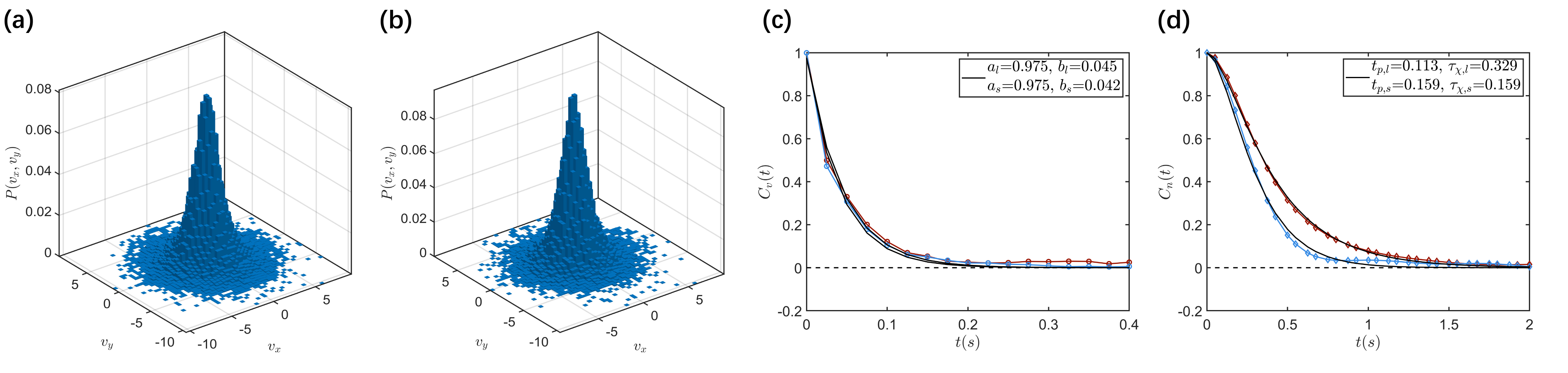}
    \caption{\label{FigE2} \color{red}\textbf{Single particle dynamics.} \textbf{(a)} {\color{red} 2D} velocity distribution $P(v_x,v_y)$ for large particles. \textbf{(b)} {\color{red} 2D} velocity distribution  $P(v_x,v_y)$ for small particles. \textbf{(c)} Single-particle velocity time correlation $C_v(t)$. \textbf{(d)} Single-particle orientation time correlation $C_n(t)$. In both panels (c) and (d), the red curve represents large particles while the blue curve represents small particles. The fitting functions are for $C_v(t)$ and $C_n(t)$ are given by a simple exponential $C_v(t)=a \exp(-t/b)$ and \eqref{fitcn} respectively. The parameters resulting from the fitting procedure are indicated as insets in panels (c) and (d).}
\end{figure*}

The competition between these effects gives rise to two distinct dynamical behaviors, observed at low and high packing fractions (panels (d) and (f) in the same Figure). 
The packing fraction is defined as the ratio of the area occupied by the particles to the total area of the system.
At low packing fractions, activity is dominant. Fig.\ref{FigE1}(d) shows the random displacement field superimposed on the individual particles within this regime, at a low packing fraction $\phi=0.309$. Due to the unique leg design, a single particle gains kinetic energy through stochastic collisions between its legs and the bottom plate, resulting in random walk motion along the {\color{red} 2D} horizontal plane. This mechanism acts independently on each particle, giving rise to random gas-like motion as illustrated in Fig.\ref{FigE1}(c). 

%In this dilute limit, the dynamics are uncorrelated, and a single particle's translational and rotational motions follow Gaussian distributions with zero mean, as demonstrated in previous studies \cite{PhysRevE.106.L052903}. Thus, these particles can be considered a granular realization of \textit{active Brownian particles} \cite{cates2015motility}.
On the other hand, as shown in Fig.\ref{FigE1}(e)), inelastic interactions tend to correlate the dynamics of the individual particles, leading to collective motion involving clusters of neighbor particles. This effect becomes more pronounced by increasing the packing fraction. In the sufficiently dense regime, particle-particle interactions dominate over individual particle activity, giving rise to collective motion. At a high packing fraction ($\phi=0.822$), large collective structures, with vortex-like form, appear in the particle displacement field, as illustrated in Fig.\ref{FigE1}(f).

In summary, the competition between activity and inelastic interactions appears to be the physical mechanism behind the emergence of a gas-like to liquid-like crossover upon increase the packing fraction \cite{jiang2024experimentalobservationgappedshear}.

\color{red}In order to precisely characterize our active system, we begin with the analysis of single-particle dynamics. As shown in Fig.~\ref{FigE2}(a)–(b), the {\color{red} 2D} velocity distribution is centered at the origin, $(v_x,v_y)=(0,0)$, for both large and small particles. This indicates that the system lacks a well-defined intrinsic velocity scale.

Furthermore, Fig.~\ref{FigE2}(c) shows the velocity time-correlation function $C_v(t)$ (see \textit{Methods}), which decays rapidly and monotonically, well described by an exponential law $C_v \propto \exp(-t/b)$ with $b=0.045$ s and $0.042$ s for large and small particles, respectively. This demonstrates that velocity correlations are finite and short-lived, with translational dynamics fully decorrelating beyond $\sim 10^{-1}$ s.

Inspired by recent theoretical work \cite{Caprini2023}, we also analyze the orientation correlation $C_n(t)$ (see \textit{Methods}), shown in Fig.\ref{FigE2}(d). Following Ref.~\cite{Caprini2023}, we fit the data to
\begin{equation}\label{fitcn}
C_n(t) = \frac{2 D_\chi t_p}{t_p^2 - \tau_\chi^2} \left(t_p \, e^{-t/t_p} - \tau_\chi \, e^{-t/\tau_\chi}\right),
\end{equation}
where $D_\chi$ is the inertial rotational diffusivity, $\tau_\chi$ the rotational memory time, and $t_p$ a characteristic timescale coinciding with the persistence time in overdamped systems. From the unit normalization of the orientation vector, $D_\chi$ can be obtained as \cite{Caprini-Lowen2022}:
\begin{equation}
D_\chi = \frac{t_p + \tau_\chi}{2\tau}.
\end{equation}

Fitting our data yields $t_{p,l} = 0.113\,\mathrm{s}$ and $\tau_{\chi,l} = 0.329\,\mathrm{s}$ for large particles, and $t_{p,s} = 0.159\,\mathrm{s}$ and $\tau_{\chi,s} = 0.159\,\mathrm{s}$ for small particles. For small particles, the fit agrees reasonably well with theory, though the tail exhibits deviations: the experimental data show an oscillatory trend absent in the theoretical curve.

Taken together, the results of Fig.~\ref{FigE2} demonstrate that our system closely resembles active Ornstein–Uhlenbeck particles (AOUPs): it exhibits a finite persistence time, but, unlike active Brownian particles, lacks a characteristic nonzero velocity scale\cite{Caprini2023,Caprini-Lowen2022}.

Moreover, we analyze the effective P\'eclet number of single particles, defined as the ratio of the persistence length to the particle size \cite{Lowen_PRL}:
\begin{equation}
\mathrm{Pe} = \frac{v_0 \tau_p}{D},
\end{equation}
where $v_0$ is the active velocity, $\tau_p$ the persistence time, and $D$ the particle diameter. We obtain 
\(\mathrm{Pe}_l = 0.0282\) for large particles and \(\mathrm{Pe}_s = 0.0358\) for small particles. The persistence time is \(\tau_p = 0.025\,\mathrm{s}\) for both particle types, 
which is much shorter than the typical observation times employed in our experiments 
\((\tau = 0.05\,\mathrm{s}, 10\,\mathrm{s}, \text{and } 500\,\mathrm{s})\) (see SI for detailed measurement).  Moreover, it is significantly smaller than the relaxation timescales shown in Fig. S5 in the Supplementary Information.

\color{black}

\begin{figure*}[htbp]
   \centering
    \includegraphics[width=1\linewidth]{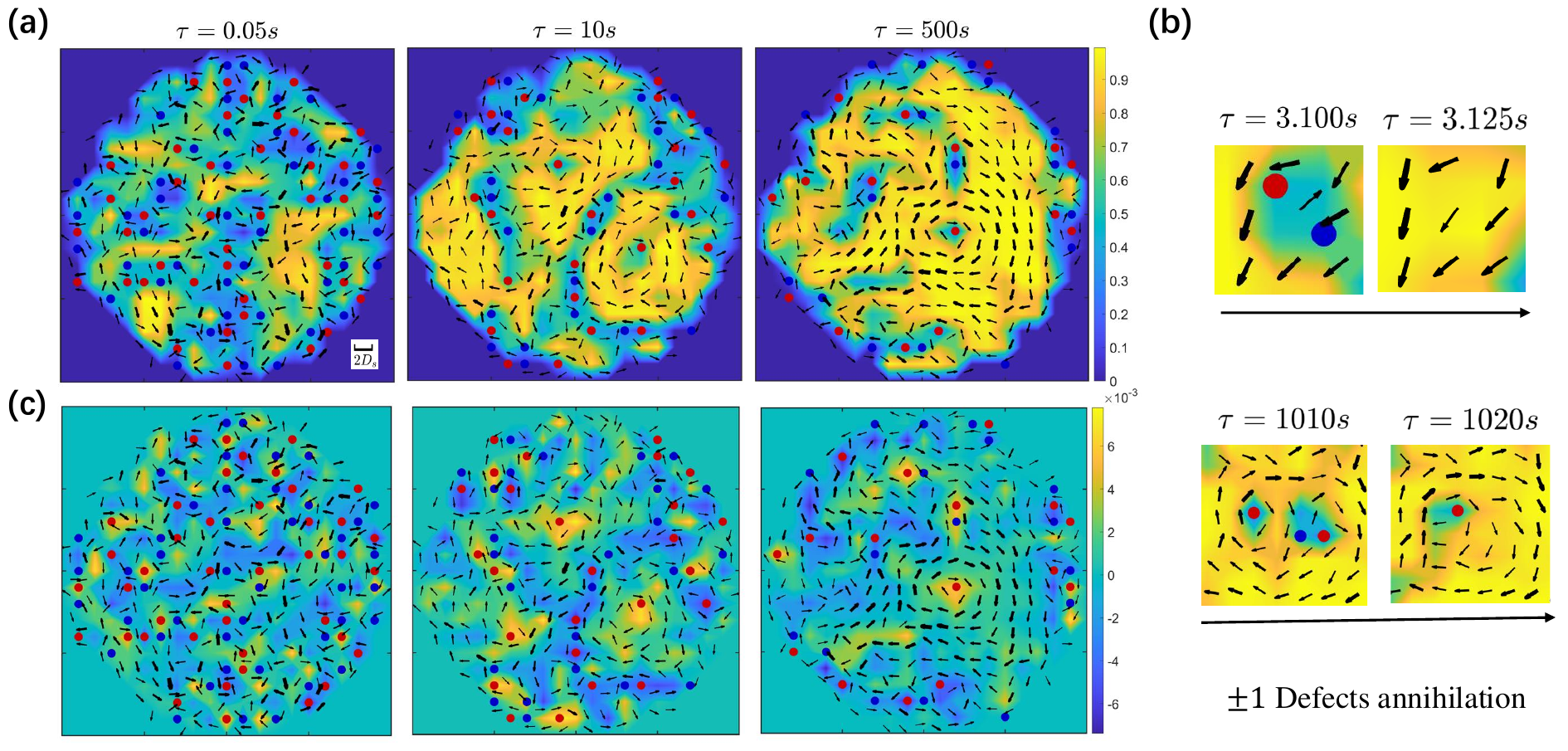}
    \caption{\label{FigE3} \textbf{Topological defects characterize the self-organization and the onset of collective motion {\color{red} at the packing fraction $\mathbf{\phi =0.822}$}.} \textbf{(a)} Displacement field (black arrows with thickness representing magnitude) for different observation times, $\tau=0.05\,s$, $10\,s$ and $500\,s$. Red and blue symbols refer to vortices and anti-vortices with $\pm 1$ winding number $q$. The color map represents the local value of the polarization $P_c$. As a reference, the white bar shows the length-scale $2 \,D_s$, with $D_s$ the diameter of the small particles. \textbf{(b)} Snapshots of two consecutive displacement frames highlighting the annihilation of a $+1$ vortex (\color{red}red\color{black}) with a $-1$ anti-vortex (\color{blue}blue\color{black}), conserving the \color{red}local topological charge \color{black} ($0$ in the top panel and $+1$ in the bottom one). \textbf{(c)} Same representation as in panel (a) but with the color map indicating the local value of the vorticity intensity $\Omega_l$.}
\end{figure*}

\section*{Self-organization and onset of collective motion}

{\color{red} In this work, unless stated otherwise, our analysis concerns a system of 798 bidisperse particles at a packing fraction of 0.822. Fig.S4(a) in the Supplementary Information shows the intermediate scattering function at this packing fraction, demonstrating that particle displacements remain minimal at late time without any evident relaxation observed within the experimental time window. In this system, particles are densely packed and the density is approximately uniform, as shown in Fig.\ref{FigE1}(f), forming the basis for the following analysis. } 

To facilitate the computation of relevant physical quantities characterizing self-organization and collective motion, we first discretize the raw particle displacement fields (such as those shown in Fig.~\ref{FigE1}(d,f)) by mapping them onto square lattice of size \(1.84\,D_s\). {\color{red} The coarse-grained displacement field is obtained by summing the displacements of all particles within each lattice cell. } More precisely, a raw displacement field is defined as $\textbf{u}_i\equiv \textbf{r}_i(t_0+\tau)-\textbf{r}_i(t_0)$, where \( \textbf{r}_i(t) \) denotes the position of particle \( i \) at time \( t \). Here, \( i \) indexes the particles, \( t_0 \) represents a specific reference time, and \( \tau \) is the duration of observation time. 
{\color{red} In our analysis, the time interval $\tau$, over which displacements are measured, can be varied freely from $\tau = 0.025$ s up to $\tau \sim 10^3$ s. To obtain reliable statistics, we divide the time sequence of each experimental run into several intervals of equal length $\tau$. For each interval $[t_0, t_0 + \tau]$, we compute the displacement field relative to the reference time $t_0$, and then collect ensemble-averaged statistics over at least 500 different $t_0$ for each analysis to ensure statistical reliability.}
\color{red} The advantage of using the displacement field, rather than the velocity field $\mathbf{v}\equiv \lim_{dt\to 0} d\mathbf{u}/dt$, is to probe the dynamics of the system at all timescales and to investigate the systematic effect of particle-particle interactions, something that is not possible by considering only the $dt \to 0$ regime.\color{black}

To quantify the spontaneous emergence of collective behavior and the emergence of self-organizing large structures, we define the local displacement polarization
\begin{equation} \label{eq-3}
  P_{c}=\frac{1}{n}\left|\sum_{l=1}^{n}\frac{\textbf{u}_{l}}{|\textbf{u}_{l}|}\right|,
\end{equation}
where $\textbf{u}_l$ stands for the local displacement vector on a given lattice site $l=1,2,\dots,n$ with $n=4$ for the chosen square lattice.

By construction, the parameter $P_c$ goes from $0$, when particle displacements are random and uncorrelated, to $1$, when all displacements point locally toward the same direction. In a sense, $P_c$ quantifies the degree of local flocking in the particle dynamics. If $n$ is taken to be the total number of particle $n=N$, then $P_c$ describes the onset of global flocking as usually employed in polar active systems, \textit{e.g.}, Vicsek model \cite{PhysRevLett.75.1226}.

To provide a further probe of collective behavior, we also introduce the intensity of the local curl of the displacement field,
\begin{equation} \label{eq-3}
\Omega_l=\left|\nabla\times\textbf{u}_l\right|=|\partial_{x}u_{l,y}-\partial_{y}u_{l,x}|,
\end{equation}
where $\textbf{u}_l$ represents the local displacement vector on a given lattice site. $\Omega_l$ is a measure of local displacement curl. A large value of $\Omega_l$ indicates a significant degree of local rotation in the displacement vector, which acts in direct opposition to the polarization $P_c$. Heuristically, one would expect $P_c$ and $\Omega_l$ to create a complementary network. This combination offers a more robust characterization of the onset of collective behavior at a local scale.

Finally, we analyze the local structure of the displacement vector field $\textbf{u}_l$ by looking for its singularities in the form of vortices and anti-vortices, \textit{i.e.}, topological defects in the dynamical displacement field. In this respect, we define the topological winding number:
\begin{equation}
    q\equiv \frac{1}{2\pi}\oint_{\mathcal{C}} d\theta,
\end{equation}
where $\theta$ is the direction (phase) of the local displacement vector concerning the $x$-axes and $\mathcal{C}$ a closed loop. $q=\pm 1$ corresponds respectively to a vortex/anti-vortex where the phase of the displacement vector is not well-defined, and the vector itself is singular.
{\color{red} It should be noted that the defects discussed here are not conventional solid-state defects (such as dislocations or Eshelby quadrupoles), but rather orientational defects, with the displacement field acting as the relevant vectorial order parameter.}

\color{red} For systems with a well-defined order parameter and stiffness, such as the XY model or superfluids, the total topological charge $q$ is conserved. In contrast, in fluid systems lacking rigidity and subject to external forces or activity, the conservation of the total topological charge is no longer guaranteed. Nevertheless, energetic arguments suggest that local conservation of $q$ is still favored, so that defects typically appear or disappear in pairs, annihilating with oppositely charged partners. It is also important to note that the very definition of topological defects requires the existence of a continuous vector field. This assumption inevitably breaks down at low packing fractions, where the particle distribution is too dilute to be well approximated by a continuous field. In what follows, we revisit these points in detail and complement them with a direct analysis of the experimental data.\color{black}

\begin{figure*}[htbp]
   \centering
    \includegraphics[width=1\linewidth]{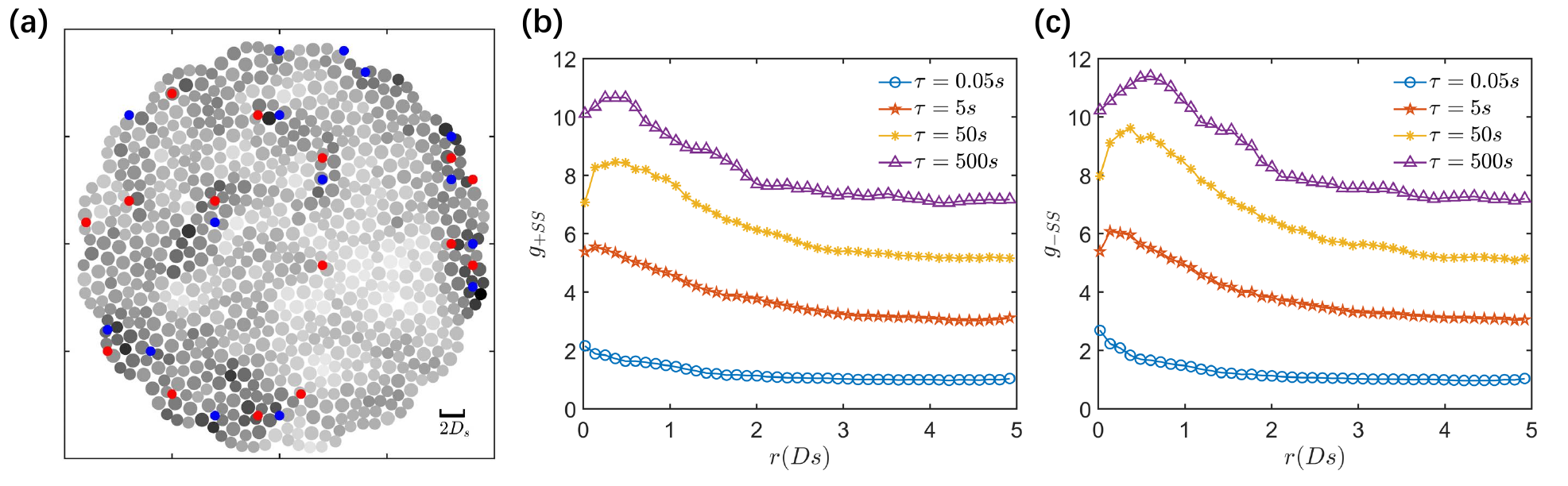}
    \caption{\label{FigE4} \textbf{Structural relaxation towards the onset of collective motion is \color{red} associated with \color{black} topological defects.}
    {\color{red} The system packing fraction is fixed at $\phi=0.822$, identical to that in Fig.\ref{FigE3}.}
    \textbf{(a)} Snapshot of $D_{\text{min}}^2$ field (gray color map) with topological defects (red and blue symbols) at the observation time of $\tau=500$ s. The characterization of local collective motion for the same $\tau$ is provided by the third panel in Fig. \ref{FigE3}(a). {\color{red}The bar indicates a length scale of $2 \,D_s$, with $D_s$ the diameter of the small particles.} \textbf{(b)} Pair correlation functions between areas of strong structural relaxation, soft spots (SS) (top $5\%$ $D_{\text{min}}^2$ particles), and positive topological defects at time interval $\tau$. \textbf{(c)} The same correlation functions for negative topological defects. Each curve has been shifted up by two units on the vertical axis for clearer visualization, {\color{red} and the horizontal axis is scaled by $\,D_s$ in (b) and (c).}}
\end{figure*}

In Fig.\ref{FigE3}, we present three snapshots of the particle displacements on the square lattice for three different observation times: $\tau=0.05\,s$, $\tau=10\,s$, and $\tau=500\,s$. The local displacement vectors are visualized with black arrows whose thicknesses represent their amplitudes. These different timescales are taken as benchmark values to characterize our system's short-, intermediate-, and long-term dynamics.
{\color{red}The snapshots in Fig.\ref{FigE3} are taken at the initial time $t_0 = 0.025 \,\text{s}$, which corresponds to the temporal resolution limit. The files \texttt{Movie S1.mp4} and \texttt{Movie S2.mp4} available in the SI \texttt{Movies} show the dynamics composed of snapshots obtained by varying the initial time $t_0$, corresponding to the observation times $\tau = 0.05 \,\text{s}$ and $\tau = 10 \,\text{s}$, respectively.}

Looking at the spatial structure of the displacements, it is clear that at short time scales ($\tau=0.05$ s in Fig.\ref{FigE3}), the motion of the particles is uncorrelated, random and independent, as in a gas-like environment. Kinematically, this is the consequence of the activity injecting kinetic energy at the individual particle level and favoring a {\color{red} 2D} \color{red}collisional motion\color{black}. Observing the dynamics on longer time scales makes the emergence of collective motion and large-scale cooperative structures evident. As we argue below, this emergence can be ascribed to the inelastic interparticle collisions and rationalized by topological concepts applied to the displacement vector field.

In Fig.\ref{FigE3}(a), we show the value of the local displacement polarization $P_c$ using a color map ranging from blue ($P_c=0$) to yellow ($P_c=1$). Thus, yellow regions represent clusters with strong collective motion, where the particle displacements are aligned as flocks of birds. As the observation time increases, from the left to right panels, the size of the cooperative (yellow) regions grows, indicating the emergence of larger-scale collective motion. To provide a microscopic explanation of this behavior, we present the location of the topological defects (TDs) in the same panels, using red symbols for vortices and blue symbols for anti-vortices. We notice that at short times ($\tau=0.05$ s), the topological defects are located in random positions, forming a gas of defects. Moreover, TDs tend to be located in regions with low polarization, indicating that TDs are microscopic objects that disrupt the onset of collective motion by randomizing the direction of the particle displacements. More details on the spatial correlation between TDs as a function of the observation time $\tau$ can be found in the Supplementary Information.

Importantly, we observe that the onset of collective motion, characterized by the percolation of clusters with large $P_c$, is mediated by the annihilation of topological defects that `live' at the interfaces between these collective clusters (blue regions). This annihilation occurs between pairs of defects with opposite charges, preserving the \color{red}local winding number\color{black}. Two examples of this process are shown with enlarged views in Fig.\ref{FigE3}(b). It is also evident and discussed in more detail below that the number of TDs decreases with the observation time, favoring the establishment of large-scale collective clusters.

{\color{red} Collective motion provides another manifestation of dynamical heterogeneity. In Fig. S4(b) of the Supplementary Information, we compute the four-point susceptibility $\chi_4(t)$ to track its evolution over the observation time. The absence of a peak in $\chi_4(t)$ within the experimental run suggests that the characteristic timescale of these heterogeneities is at least several thousand seconds—much longer than the observation window considered in this study.}

{\color{red} Previous work by Silke Henkes and colleagues also demonstrated large-scale correlated dynamics reminiscent of dynamical heterogeneities \cite{Henkes_NC,Henkes_PRE}. However, and crucially different from our case, their system corresponds to a jammed solid phase, whose dynamics are dominated by elasticity.}

To provide a complementary view of these dynamics, in Fig.\ref{FigE3}(c), we present the same particle displacements and TDs but, this time, superimposed onto the amplitude of the local displacement curl $\Omega_l$. As expected, the local curl color map (Fig.\ref{FigE3}(b)) and the local polarization color map (Fig.\ref{FigE3}(a)) are roughly complementary. Moreover, the local curl $\Omega_l$ is significant around the TDs, that can therefore be identified as the microscopic objects disrupting the irrotational nature of the displacement field. In this complementary view, the onset of collective motion is driven by the percolation of areas with low local displacement curl, {\color{red} which is associated with} the annihilation of defect pairs (Fig.\ref{FigE3}(c)).

\begin{figure*}[htbp]
   \centering
    \includegraphics[width=1\linewidth]{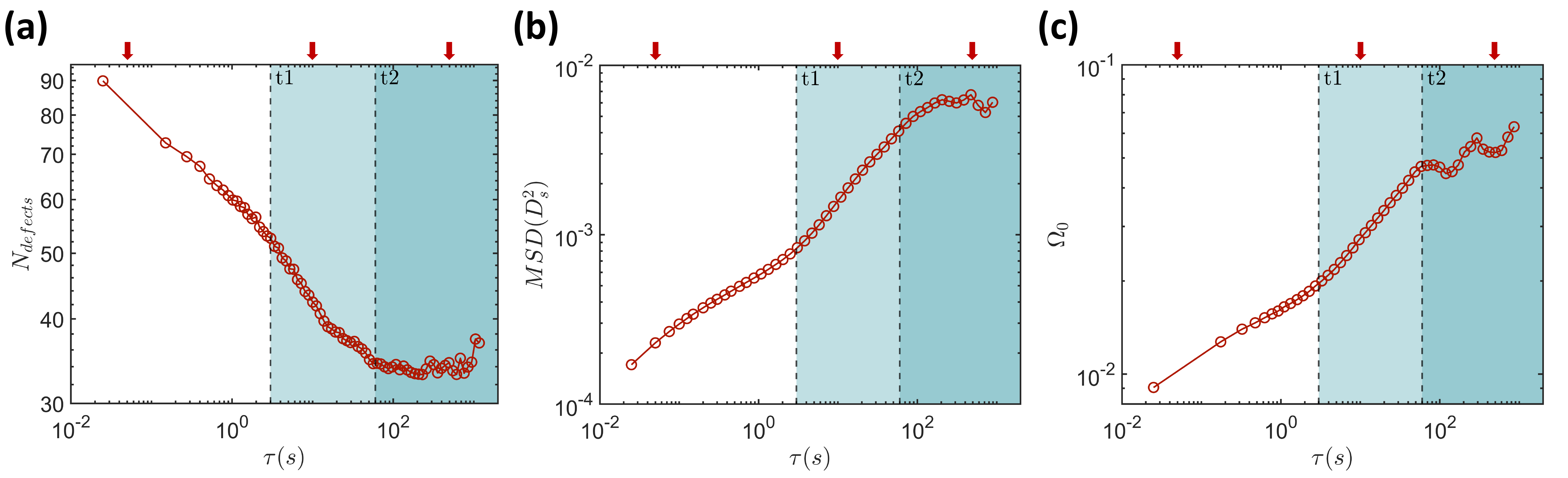}
    \caption{\label{FigE5} \textbf{Three stage non-equilibrium dynamics {\color{red} at the packing fraction of 0.822}.} \textbf{(a): \textbf{Topological signatures}.} Defect statistics as a function of the observation time $\tau$. $N_{\text{defects}}$ is the total number of TDs. The vertical dashed lines indicate the two time scales $t_1=3\,s$, $t_2=60\,s$. \textbf{(b): Particle-level dynamics.} Mean square displacement (MSD) in units of the square of the small particle diameter {\color{red} $D_s^2$}. \textbf{(c): Collective motion.} Average curl of the system $\Omega_0$. The three red arrows at each panel correspond to the selected times shown in Fig.\ref{FigE3},
    {\color{red}while the colored backgrounds indicate the system’s dynamics across the three stages.} }.
\end{figure*}

\section*{Relaxation towards a non-equilibrium steady state is mediated by topological defects
}
As discussed in the previous Section and shown in Fig. \ref{FigE3}, topological defects appear at the interface between regions with highly collective motion, and their annihilation is the microscopic mechanism behind the onset of large-scale structures in the dynamics. As a result, the relaxation towards the final non-equilibrium steady state is concentrated at the interface between these regions (\textit{e.g.}, yellow areas in Fig. \ref{FigE3}(a)).

To explore this physical mechanism further, we calculate $D_{\text{min}}^2$, which quantifies the strength of nonaffine particle displacements between two consecutive configurations (see Methods \ref{methods} for details) \cite{falk1998nonaffine}. A large value for this quantity represents soft areas in the sense that structural relaxation is more intense than elsewhere. In Fig.\ref{FigE4}(a), we calculate the intensity of $D_{\text{min}}^2$ projected on each particle for $\tau=500s$ using a gray color map, with darker regions representing larger values of $D_{\text{min}}^2$. By visually comparing this color map with that presented in the third panel of Fig.\ref{FigE3}(a), it is clear that regions with large $D_{\text{min}}^2$ correspond to smaller values of $P_c$. This implies that areas with strong structural relaxation are those where motion is still random and poorly collective.

In Fig.~\ref{FigE4}(a), we represent the topological defects (TDs) with charges \( +1 \) and \( -1 \) using blue and red symbols, respectively. These TDs appear in pairs within regions of higher \( D_{\text{min}}^2 \) values. To quantify this observation, we define `\textit{soft spots}' (SS) as the top 5\% of particles with the highest \( D_{\text{min}}^2 \) values and analyze their positional correlation with both positive and negative defects. The results are shown in Fig.~\ref{FigE4}(b) and (c).
Each curve exhibits a peak at small \( r \), with the peak height increasing slightly over time. This suggests a strong short-range correlation between soft spots and TDs. The correlation is slightly stronger for negative defects, though the difference is not substantial. This could be explained by the fact that TDs typically appear in pairs, which may balance out the correlation differences.
Overall, the observations in Fig.~\ref{FigE4} suggest that topological defects {\color{red} are involved in} the system's structural relaxation and {\color{red} contribute to} the emergence of collective motion during out-of-equilibrium dynamics.

We also notice that the strong correlation between structural relaxation ($D_{\text{min}}^2$) and topological vortex-like defects is reminiscent of recent results in 2D amorphous solids and defective crystals in which these objects have been proposed as the microscopic carriers of plasticity \cite{wu2023topology,baggioli2023topological, PhysRevE.109.L053002,bera2024clustering,vaibhav2025experimental,huang2024spotting}.

\section*{Three stage non-equilibrium dynamics}
We investigate the non-equilibrium dynamics of the system {\color{red} at $\phi=0.822$} across a full range of observation times, spanning six decades from $\tau=10^{-2}\,s$ to $\tau=10^4\,s$, and reveal a three-stage evolution that is evident in both topological characteristics, particle level dynamics, and collective motion. Three snapshots of the system dynamics in each of these stages have already been presented in Fig. \ref{FigE3} (see vertical red arrows in Fig. \ref{FigE5}). \color{red}An alternative representation highlighting the crossover between the first and second regimes is provided in the Supplementary Information\color{black}.

In Fig.~\ref{FigE5}(a), we track the evolution of topological defects at different time scales. Up to a characteristic time scale of $t_1 \approx 3\,s$, the total number of defects, $N_{\text{defects}}$, decreases monotonically over time, indicating frequent and intermittent annihilation events. Beyond $t_1$, coinciding with the strengthening of collective motion shown in Fig.~\ref{FigE3}, the dynamics of TDs accelerate as the average collective cluster size increases. Consequently, the total number of defects declines at a faster rate.  
At a second characteristic time scale, $t_2 \approx 60\,s$, the system reaches a non-equilibrium steady state, likely due to a balance between activity and inelastic interactions. Beyond this point, the number of topological defects remains relatively constant. Simultaneously, large vortex clusters begin to percolate through the space, culminating in a dominant vortex-like structure particularly visible at the center of the third panel in Fig.~\ref{FigE3}(a).

Figure~\ref{FigE5}(b) displays the mean square displacement (MSD), where the unit length is given in terms of the small particle diameter ({\color{red} $D_s$}). This quantity characterizes the system's dynamics at the particle level. The MSD evolution follows three distinct stages.  
For $t < t_1$, the MSD exhibits sub-diffusive behavior, $\text{MSD} \propto t^{0.32}$, indicating that particle motion is constrained within a `cage' on short timescales. Between $t_1$ and $t_2$, as the total number of TDs, $N_{\text{defects}}$, decreases more rapidly, particle dynamics accelerate while maintaining sub-diffusive characteristics with a different exponent, $\text{MSD} \sim t^{0.55}$. This regime coincides with the increasing mean cluster size of collective motion, as observed in the second snapshot of Fig.~\ref{FigE3}(a).  
Beyond $t_2$, particle dynamics slow as the system reaches a non-equilibrium steady state, where the MSD approaches a plateau. {\color{red} This indicates that at the highest packing fraction $\phi=0.822$, the system is in a dense
liquid-like state, and the particle motion is very limited, i.e., frozen.} This trend aligns with the stabilization of the total number of defects, {\color{red}consistent with defects potentially contributing to} local relaxation. Ultimately, the system attains a collective steady state, emerging from the interplay between activity (energy injected through vibrations) and inelastic collisions.

To further investigate the relationship between the formation of vortices in the system and its collective  dynamics, we calculate the average curl $\Omega_0$ of the displacement field, defined as 
\begin{equation} \label{eq-3}
  \Omega_{0}=\sqrt{\left<\frac{1}{N}\sum_{l}(\nabla\times\textbf{u}_{l})^2\right>}.
\end{equation}
As shown in Fig.\ref{FigE5}(c), this quantity also shows a characteristic three-stage evolution. It slowly increases before $t_1$, rises rapidly between $t_1$ and $t_2$, and reaches a steady state after $t_2$. The average curl behavior is consistent with the defect numbers and MSD, confirming the previous physical picture.

%{\color{red} It is interesting to relate our two-stage relaxation, with its first and second stages starting at $t_1$ and $t_2$, to the conventional time scales observed in glassy systems. However, our experimental time window remains considerably shorter than the typical time scales required to fully address such issues \cite{Berthier_RMP}. Nevertheless, by comparing Fig.\ref{FigE5} in our manuscript with Fig.~10 in Berthier and Biroli’s work \cite{Berthier_RMP}, we conjecture that the two-stage relaxation observed in Fig.\ref{FigE5} may correspond to the microscopic relaxation and the early $\beta$-relaxation processes.}

In conclusion, the number of defects, MSD, and average curl exhibit three distinct dynamical stages, establishing a link between topology, particle-level dynamics, and collective motion. These quantities are interrelated: at short time scales ($t < t_1$), particle motion is ``trapped'' due to frequent collisions, resulting in slow sub-diffusion in the MSD. Rapid fluctuations in the displacement field, driven by collisions, lead to efficient defect annihilation, causing a continuous decrease in defect numbers.  
Between $t_1$ and $t_2$, the system's dynamics evolve more rapidly, driven by the formation of larger vortices in the displacement field. During this period, the defect number decreases at an accelerated rate.  
After $t_2$, the system reaches a steady state, where the number of defects, MSD, and average curl stabilize over time. The average curl attains its maximum value, signifying the emergence of stable large vortices. This accounts for the total topological number approaching a plateau and the onset of large-scale collective motion. 
%Defects regulate local structural rearrangements, ultimately culminating in the emergence of large vortices over long timescales.
{\color{red}In our system, defects are associated with local structural rearrangements that can contribute to the emergence of large vortices over long timescales. By analyzing the statistics of these topological defects, we obtain a simple and computationally inexpensive measure that captures the system’s three-stage dynamical behavior (see Fig.~\ref{FigE5}), showing excellent agreement with more traditional dynamical and coarse-grained observables.
}

\begin{figure}
   \centering
    \includegraphics[width=\linewidth]{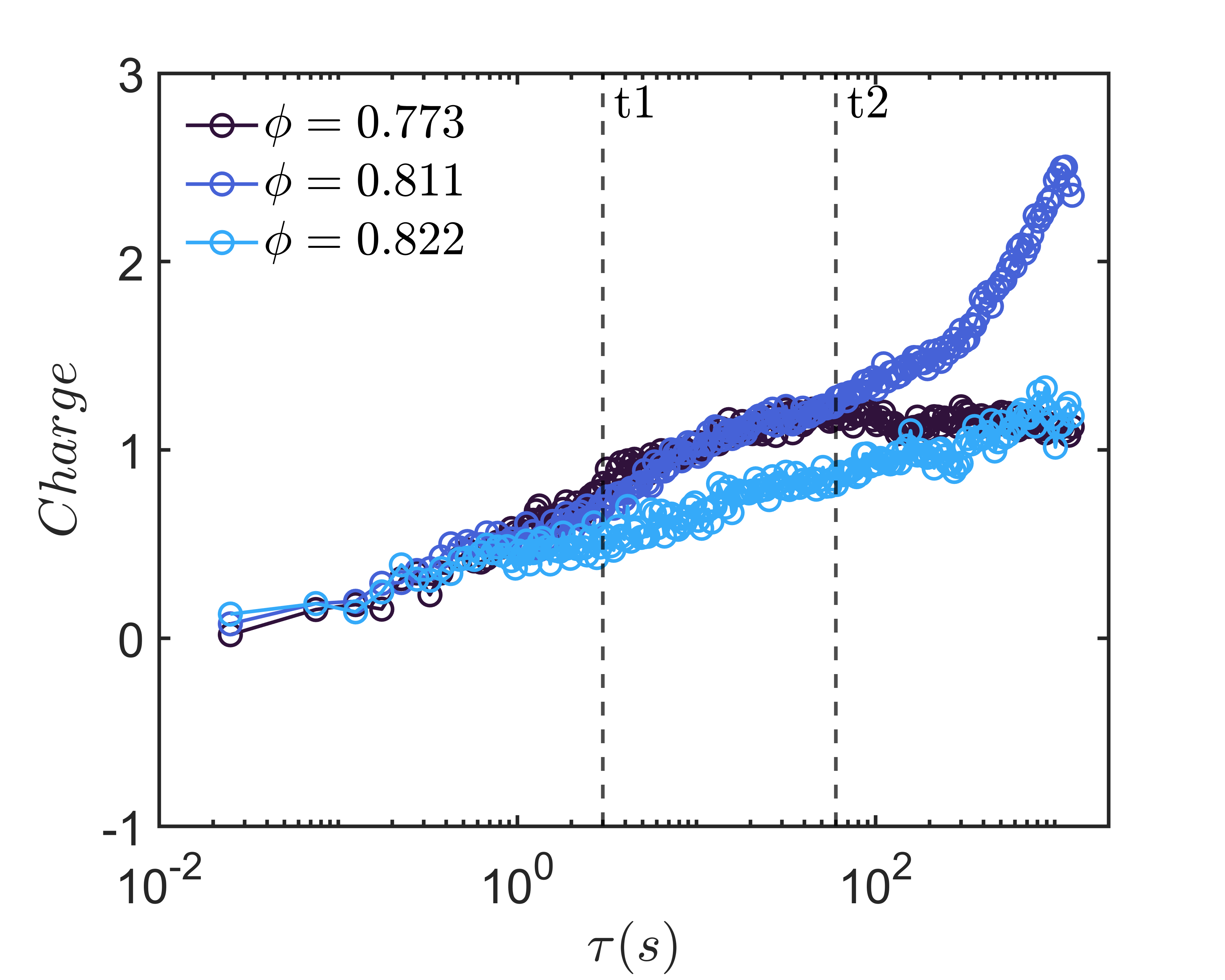}
    \caption{\color{red}\label{FigE6} \textbf{Evolution of the average total topological charge with observation time $\mathbf{\tau}$.} Different colors correspond to $\phi=0.773,0.811,0.822$. The vertical dashed lines indicates the characteristic timescales $t_1,t_2$ for the highest packing fraction $\phi=0.822$.}
\end{figure}

\color{red}To probe more deeply the topological nature of the active dynamics, we track the total topological charge, defined as the sum of the winding numbers of all defects in a single snapshot. In rigid systems with a well-defined order parameter and controlled boundaries, this charge is conserved in time. By contrast, in non-rigid fluids subject to activity and uncontrolled boundaries, the situation is more subtle.

We analyze three packing fractions, $\phi=0.822, 0.811, 0.773$, and compute the average total charge as a function of the observation timescale $\tau$. The results, averaged over $500$ realizations, are shown in Fig.~\ref{FigE6}.

At short times, below a critical scale $t^* \approx 10^{-1}$ s, the total charge remains approximately zero and conserved. This regime coincides with the solid-like behavior of our dense active fluid, previously linked to propagating shear waves and high-frequency rigidity \cite{jiang2024experimentalobservationgappedshear}.

At longer times, $t > t^*$, the system loses rigidity and charge conservation breaks down. For $\phi=0.822$ and $\phi=0.773$, the average charge stabilizes near unity at late times—remarkably close to the Euler characteristic of a {\color{red} 2D} disk, which via the Poincaré–Hopf theorem fixes the total winding number to one. Since our system departs from the theorem’s assumptions (boundary conditions, activity, dissipation), whether this agreement is coincidental or physically meaningful remains to be understood.

By contrast, $\phi=0.811$ shows a steady growth of the average charge with time, pointing to a more intricate interplay between topology, activity, and dissipation, an avenue that calls for further exploration.
\color{black}

\section*{Turbulent-like energy cascades}
The non-equilibrium characteristics of vortices offer a microscopic perspective on the system's dynamics and the emergence of large-scale collective motion in the final non-equilibrium steady state. Kinematically, the transition to this steady state is governed by the interplay between activity and inelastic collisions.  
As energy is injected at short scales (\textit{i.e.}, the particle level), the emergence of self-organized collective motion necessitates an energy transfer from small to large scales, a hallmark of turbulent-like inverse cascades.  
In fluid mechanics, the inverse energy cascade was first introduced to describe the spatial-temporal chaotic flow dynamics in 2D turbulence, considering the conservation of kinetic energy and mean squared vorticity in the inertial range where energy dissipation is negligible. This phenomenon was first predicted in the late 1960's by Kraichnan \cite{10.1063/1.1762301} and later confirmed by numerous studies (\textit{e.g.}, \cite{10.1063/1.863225,10.1063/1.864870,PhysRevLett.81.2244,PhysRevLett.85.976, PhysRevLett-Tabeling}). A notable natural example of this process is the formation of massive quasi-2D structures, such as hurricanes.

{\color{red}In fluid turbulence, the inverse cascade is typically characterized in the inertial range through the third-order spatial correlation of velocity fluctuations, which reveals both the cascade process and its direction. For macroscopic active granular particles, however, the absence of a clear inertial range and the strongly dissipative nature of inelastic collisions and friction render this approach ineffective, as energy loss far exceeds any flux transferred through a cascade. We therefore adopt a different strategy: rather than attempting to quantify the exact flux, we focus on determining the \textit{direction} of the cascade, which is experimentally more accessible.}

\begin{figure*}[htbp]
   \centering
    \includegraphics[width=1\linewidth]{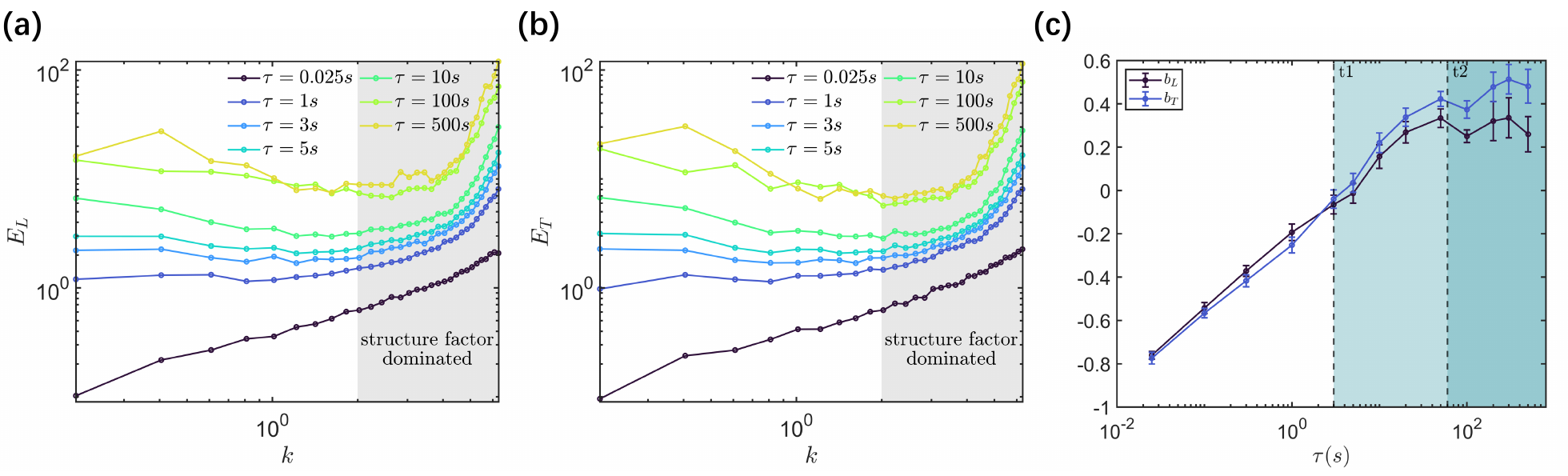}
    \caption{\label{FigE7} \textbf{Turbulent-like energy cascades as revealed by energy power spectra {\color{red} at $\phi=0.822$}.} Longitudinal \textbf{(a)} and transverse \textbf{(b)} kinetic energy spectra at different observation times $\tau$ as a function of the wave number $k$. 
    {\color{red}The gray shaded background indicates the region dominated by the structure factor.}
    \textbf{(c)} Power-law exponents $b$ for longitudinal ($b_L$) and transverse spectra ($b_T$).
    {\color{red}The background colors in panel (c) correspond to the system’s three dynamical stages shown in Fig.\ref{FigE5}.}}
\end{figure*}

To test this hypothesis, we define the kinetic energy spectra,
\begin{equation} \label{eq-3}
  E_{L}(k)\equiv2\pi{k}\left<\frac{1}{N}\left|\sum_{j=1}^{N}{u}_{j,\parallel}e^{-i \mathbf{k}\cdot\mathbf{r}_{j}}\right|^{2}\right>,
\end{equation}
\begin{equation} \label{eq-3}
  E_{T}(k)\equiv2\pi{k}\left<\frac{1}{N}\left|\sum_{j=1}^{N}{u}_{j,\perp}e^{-i\mathbf{k}\cdot\mathbf{r}_{j}}\right|^{2}\right>,  
\end{equation}
where $N$ is the number of particles in the system. ${u}_{j,\parallel}$ and ${u}_{j,\perp}$ are the components of the displacement vector of a particle $j$ parallel and perpendicular concerning the wave vector $\mathbf{k}$ and $i$ is the imaginary unit. $L$ and $T$ stand respectively for ``longitudinal`` and ``transverse''.

\begin{figure*}[htbp]
   \centering
    \includegraphics[width=0.9\linewidth]{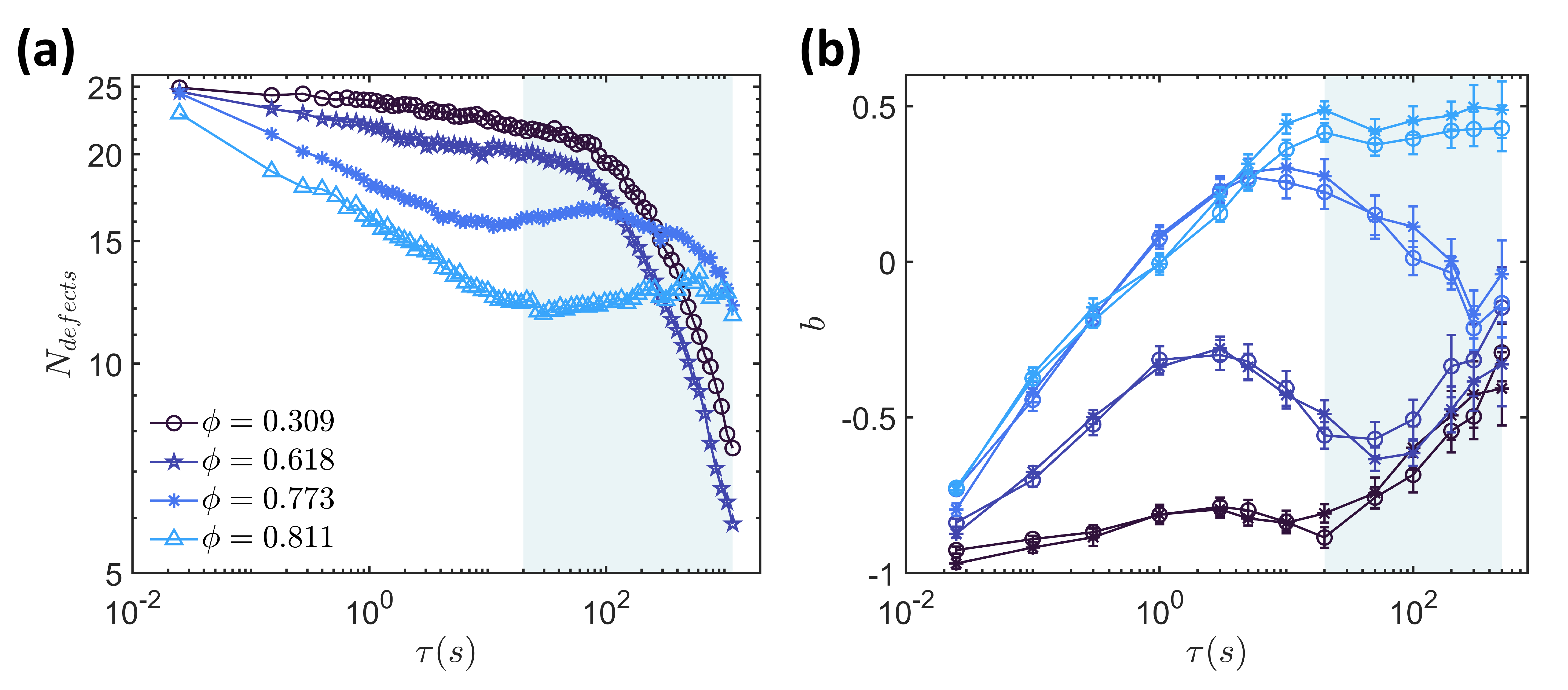}
    \caption{\label{FigE8} \textbf{Activity versus interactions and non-equilibrium steady state formation.} \textbf{(a): } Defect statistics as a function of the observation time $\tau$ for different packing fractions $\phi$. All previous figures refer to the highest packing fraction $\phi=0.822$. $N_{\text{defects}}$ is the total number of TDs. \textbf{(b): } Power-law exponents $b$ for energy spectra. The color scheme refers to the different packing fractions shown in panel (a). Open circles denote the longitudinal component $b_L$ and stars the transverse component $b_T$. The background color in both panels indicate the onset of the non-equilibrium steady state at high packing fraction, to be contrasted with its absence at low $\phi$.}
\end{figure*}

Our experimental results are presented in Fig.~\ref{FigE7}(a, b) at various time scales, denoted as \(\tau\). At large wave vectors—or equivalently, very short length scales—the curves are strongly affected by the static structure factor, which arises from the hard-disk interactions in our granular system. This factor typically exhibits a peak around \( k \approx 2\pi \), though it is not displayed in the figure.  
It is important to note that the unit of length is expressed in terms of the small particle diameter {\color{red}\( D_s \)}. Moreover, as the observation time \( \tau \) increases, the influence of the static structure factor becomes more pronounced. This occurs because the spectra \( E_L(k) \) and \( E_T(k) \) are computed using the displacement field rather than the velocity field. Consequently, this would introduce a \( \tau \) dependent scaling prefactor for \( E_L(k) \) and \( E_T(k) \), revealing that kinetic energy fluctuations diminish as a function of \( \tau \).  
As a result, at larger observation times, the effect of the static structure factor becomes increasingly evident around particle scales.

Nevertheless, a clear trend emerges in the intermediate and low range of \( k \). In particular, at short times, both $E_L$ and $E_T$ grow with $k$, flattening in an intermediate $\tau$ range and then changing their slope at late times. To better characterize this behavior, we fit the various curves with a power-law ansatz $E_{L,T} \propto k^{-b_{L,T}}$. The extracted power-law exponent is presented in Fig. \ref{FigE7}(c) as a function of the observation time $\tau$. Below $t_1$, we observe that $b_{L,T}$ are approximately identical and both negative, {\color{red}reflecting a spectrum dominated by small spatial scales, consistent with energy injection at the particle level by external vibrations.}
Around $t_1$, the spectrum starts to flatten, and the power-law exponents become slightly positive, presenting differences between the longitudinal and transverse sectors.
{\color{red} Here $b_{L,T}=0$ denotes a flat spectrum with equal energy across different length scales, while $b_{L,T}>0$ signals energy transfer toward larger spatial scales. The evolution of $b_{L,T}$ with $\tau$ thus provides a clear indication of the cascade direction.}
Finally, above $t_2$, the power-law exponents reach a plateau value in approximately $b_L\approx 0.3$ and $b_T \approx 0.4$.

{\color{red} While Henkes \textit{et al.} predicted 
$E(q) \sim q^{-1} \quad (b = 1)$ \cite{Henkes_NC},
our measurements yield distinct exponents. This discrepancy can be attributed to the fact that, even at the highest packing fraction studied (\(\phi = 0.822\)), the system remains in a liquid state rather than a jammed solid \cite{jiang2024experimentalobservationgappedshear}.}

This kinematic behavior aligns well with the three-stage dynamics presented in Fig.\ref{FigE5} and can be rationalized as follows. At short time scales, the energy injected by the active driving is gradually transferred to the individual particles, inducing a direct energy cascade towards microscopic scales. Around $t_1 \approx 3\,s$, energy is evenly distributed across all scales, {\color{red} as indicated by the scaling exponent $b\approx 0$ shown in Fig.\ref{FigE8}(b).} When the observation time exceeds $3\,s$, energy starts transferring from small to larger scales, exhibiting a turbulent-like inverse cascade. This process becomes steady around $t_2\approx 60\,s$ when the energy transfer becomes roughly independent of time.

The results in Fig.~\ref{FigE7} reveal intriguing scaling behaviors, prompting further investigation into the underlying physical mechanisms. A natural question arises: what physical mechanism is responsible for this behavior? 
In 2D turbulence, the theory of inverse energy cascade assumes the conservation of mean kinetic energy and mean squared vorticity in the inertial range~\cite{10.1063/1.1762301}. In our active granular system, not all kinetic energy is transferred from small to large scales due to strong dissipation caused by inelastic collisions and friction. Nevertheless, as we will argue below, a similar principle governs the self-organization dynamics in our system, leading to a turbulent-like inverse energy cascade. 

In Fig.~\ref{FigE5}(c), the mean curl of the displacement field, $\Omega_0$, exhibits two distinct power-law scaling behaviors in the first two regimes when $\tau<t_2$, following $\Omega_0 \propto \tau^{\alpha}$ with $\alpha=0.15$ for $\tau<t_1$ and $\alpha=0.3$ for $t_1<\tau<t_2$. The quantity $\Omega_0$, which characterizes the displacement field $\textbf{u}$ over an observation time $\tau$, differs from vorticity in 2D turbulence in two key aspects. 
First, in turbulence, vorticity is derived from the velocity field, $\Omega \equiv \nabla \times \textbf{v}$, rather than the displacement field. In a fluid system, the two quantities are directly proportional only at sufficiently short observation times. However, in a granular system, the cascade process is further complicated by inelastic collisions and friction, which prevent full energy transfer across scales. Second, the displacement field $\textbf{u}$ exhibits a nontrivial dependence on $\tau$. 

%We assume that the mean squared vorticity, $\overline {\Omega^2}$, follows $\overline {\Omega^2} \propto (\Omega_0/ \tau^{\alpha})^2$. By definition, this implies $\overline {\Omega^2} \propto \tau^{-2\alpha} \textbf{u}_l^2$ from dimensional analysis. Estimating $\textbf{u}_l^2$ from the MSD results in Fig.~\ref{FigE5}(b), we obtain $\textbf{u}_l^2 \propto \tau^{\beta}$, with $\beta=0.3$ for $\tau<t_1$ and $\beta=0.55$ for $t_1<\tau<t_2$. This leads to $\overline {\Omega^2} \propto \tau^{\beta-2\alpha}$. Notably, the exponent $\beta-2\alpha \approx 0$ for both $\tau<t_1$ and $t_1<\tau<t_2$. These findings suggest that the mean squared vorticity is nearly conserved in the turbulent-like inverse energy cascade of our active granular system, adhering to a principle akin to that of 2D turbulence.

{\color{red} To provide a self-consistent check of our independent measurement, we assumed that the mean squared vorticity (also known as \textit{enstrophy}), 
$\overline{\Omega^2}$, follows
$\overline{\Omega^2} \propto \left( \frac{\Omega_0}{\tau^{\alpha}} \right)^2$.
This assumption draws an analogy with the turbulence cascade in 2D, where $\overline{\Omega^2}$ is conserved. 
Moreover, since the mean curl of the displacement field scales as 
$\Omega_0 \propto \tau^{\alpha}$,
this assumption removes the explicit time dependence from $\overline{\Omega^2}$.

We then performed a self-consistent check to determine whether this assumption aligns with the scaling of the 
displacement field, which itself depends nontrivially on $\tau$, namely ${\bf u}^2 \propto \tau^\beta$. This time dependence can be estimated independently from the single-particle MSD curves in Fig.~\ref{FigE5}(b). 
Finally, we find that the results are consistent, as
$\beta - 2\alpha \approx 0 $
for both $\tau < t_1$ and $t_1 < \tau < t_2$. 

From a physical perspective, this supports our initial assumption that the mean squared vorticity (enstrophy) is conserved, as expected for traditional two-dimensional turbulent flows.
}

\section*{Packing-fraction controlled interplay between activity and interactions}

We conducted experiments at various packing fractions to gain deeper insights into the collective behavior of the system and the competition between activity and inelastic collisions (see Fig. \ref{FigE1}).

Our initial analysis focused on the highest packing fraction ($\phi=0.822$), where strong inelastic collisions balance the effects of activity, resulting in a late-time non-equilibrium steady state. Here, we extend the analysis to the lowest packing fraction ($\phi=0.309$), where the dynamics remain decorrelated and gas-like due to the weak inelastic interactions \cite{jiang2024experimentalobservationgappedshear}.
{\color{red} For low packing fraction, $\phi = 0.309$, the system contains many empty domains. In this regime, the topological description is not well defined since the displacement vector cannot be reliably approximated by a continuous field. Nevertheless, for completeness, we also present the results for $\phi = 0.309$ in Fig. \ref{FigE8}(a).}

Fig.\ref{FigE8}(a) shows the evolution of the total number of topological defects, $N_{\text{defects}}$, for four packing fractions ($\phi=0.309$, $0.618$, $0.773$, and $0.811$). At a high packing fraction ($\phi=0.811$), the behavior of $N_{\text{defects}}$ resembles that observed at the maximal packing fraction shown in Fig. \ref{FigE5}(a).

The instability of the non-equilibrium steady state becomes evident at $\phi=0.773$, where the system only forms a transient plateau before decaying at late times. Notably, this late-time decay is significantly faster than the initial short-time decay, a pattern also observed at the two lower packing fractions.

At the lower packing fractions ($\phi=0.309$ and $\phi=0.618$), the number of defects decreases monotonically, exhibiting a two-step relaxation mechanism without ever reaching a non-equilibrium steady state. This can be attributed to the diminishing intensity of inelastic interactions as the packing fraction decreases. Since these interactions are crucial for establishing collective motion by balancing the random effects of activity (see Fig. \ref{FigE1}), their weakness in these dilute systems prevents the formation of a steady state, as reflected by the continuous decline in the number of defects.

To validate this interpretation, Fig. \ref{FigE8}(b) presents the power-law exponents for turbulent-like energy cascades in active granular fluids across the four packing fractions. At the lower packing fractions ($\phi=0.309$ and $\phi=0.618$), the power-law exponents $b$ remain negative throughout the observation period, indicating that energy transfer from small to large scales never occurs. The absence of this inverse cascade underscores the system's inability to form a stable steady state characterized by constant values for both the longitudinal and transverse components of $b$. This finding emphasizes that the inverse energy cascade, driven by topological defect dynamics, is fundamental to establishing a non-equilibrium steady state with emergent large-scale collective motion.

For the intermediate packing fraction ($\phi=0.773$), the power-law exponents reach positive values but fail to stabilize over long timescales, indicating an increased yet insufficient attempt to achieve collective behavior. In contrast, at the high packing fraction ($\phi=0.811$), the power-law exponents continue to rise and eventually plateau over extended timescales, as shown within the blue-shaded region of Fig. \ref{FigE8}(b).

{\color{red} In Fig.\ref{FigE8}, the strongest nonmonotonic behavior appears at $\phi=0.618$ and $\phi=0.773$, both in the defect number and in the scaling exponents $b_{T,L}$. An energy spectrum with $b_{T,L}\approx0$, corresponding to nearly equal energy across length scales, is highly unusual and has previously been linked to uncorrelated bacterial swimmers in the dilute limit\cite{ChengXiang_SM,Joakim_SM}. In our system, however, this behavior originates from transient domains of collective motion that coexist with disordered regions and rapidly reorganize, as illustrated by the displacement field at $\phi=0.773$, $\tau=1\,\mathrm{s}$ (see \texttt{Movie S3.mp4} in the SI Movies). Such spatiotemporal mixtures generate flat spectra and give rise to the nonmonotonic trends in Fig. 6. This interpretation is consistent with the sharp increase of the structural relaxation time around $\phi\approx0.773$ (see Fig.S5 in the Supplementary Information), which indicates glassy dynamics.}

This comprehensive analysis highlights the interplay between activity and interactions. At low packing fractions, weak inelastic collisions allow activity to dominate, injecting energy at microscopic scales without propagating to larger ones. As a result, the system remains decorrelated and fails to form a non-equilibrium steady state. In contrast, at high packing fractions, interactions gradually balance with activity, enabling energy transfer to larger scales and the formation of collective motion. Remarkably, fully established collective motion is observed only for $\phi=0.811$ and $\phi=0.822$, aligning with previous studies that identified liquid-like collective behavior at these packing fractions \cite{jiang2024experimentalobservationgappedshear}. This balance between activity and inelastic interactions likely gives rise to liquid-like collective dynamics, suggesting that the late-time steady state behaves similarly to a thermal liquid \cite{Trachenko_2016,BAGGIOLI20201}.

\section*{Conclusions}
Using active granular vibrators as a model for \color{red} apolar dry active Ornstein–Uhlenbeck particles\color{black}, we conducted a comprehensive experimental study to investigate the topological features that give rise to steady-state large-scale collective dynamics and turbulent-like inverse energy cascades. The dynamical behavior of the system evolves through a packing fraction-controlled interplay between activity and particle interactions.

Activity at the particle level converts the injected mechanical energy into Brownian-like stochastic motions, while particle-particle interactions, driven by inelastic collisions and friction, promote coherent motion. At sufficiently high packing fractions, the effects of inelastic collisions gradually dominate over active contributions as observation time increases, leading to the emergence of large-scale collective motion.

This self-organization occurs through structural rearrangements {\color{red} associated with} topological defects, including the annihilation of positive and negative defects and the formation of large vortices. The number of defects, the mean squared displacement of the system, and the average curl of the displacement field together reveal coherent three-stage dynamics, highlighting the {\color{red}notable} role of defects in the system's evolution, {\color{red}which can serve as an efficient and practical way to quantify the system’s behavior.} From a kinematic perspective, defect annihilation and the creation of large-scale collective regions are associated with a turbulent-like inverse energy cascade, where energy transfers from small to large scales.

Our study demonstrates how large-scale collective motion emerges from a homogeneous ensemble of \color{red}apolar dry active Ornstein–Uhlenbeck particles\color{black}, offering new insights into collective dynamics in biological and robotic systems, such as insect swarms, bird flocks, and robotic swarms.

\begin{acknowledgments}
CJ and MB acknowledge the support of the Shanghai Municipal Science and Technology Major Project (Grant No.2019SHZDZX01). MB acknowledges the support of the sponsorship from the Yangyang Development Fund.
ZZ, YC and JZ acknowledge the support of the NSFC (No. 11974238 and No. 12274291) and the Shanghai Municipal Education Commission Innovation Program under No. 2021-01-07-00-02-E00138. ZZ, YC and JZ also acknowledge the support from the Shanghai Jiao Tong University Student Innovation Center.
\end{acknowledgments}

\section*{Methods}\label{methods}
%\section{Methods}\label{methods}

\subsection{Single-particle correlation function}
For single-particle dynamics (Fig.~\ref{FigE2}), the velocity autocorrelation function is defined as
\begin{equation}
C_v(t) = \left\langle \frac{\mathbf{V}(0)\cdot \mathbf{V}(t)}{\mathbf{V}(0)\cdot \mathbf{V}(0)} \right\rangle ,
\end{equation}
where $\mathbf{V}(t) = \frac{\mathbf{r}(t+\delta t)-\mathbf{r}(t)}{\delta t}$ with $\delta t = 0.025\,\mathrm{s}$. The orientation correlation function is given by
\begin{equation}
C_n(t) = \left\langle \cos[\theta(t) - \theta(0)] \right\rangle .
\end{equation}
Since the particles are non-polar, $\theta$ is defined by a short marker line on each particle to enable tracking of rotational motion (Fig.~\ref{FigE1}(a)).

\subsection{Identification of topological defects}
The topological defects are defined using the discrete displacement field. 
The vortex detection algorithm steps around the 2 × 2 square in the conventional counterclockwise sense, computing the phase gradient, \textit{i.e.}, the difference values in displacement. When the modulus of this difference $\left|\Delta\chi\right|$ exceeds $\pi$, the phase is unwrapped. Then $p$, the detected winding number is found such that the phase gradient is $2p\pi-\Delta\chi\mapsto\Delta\xi$, where $p\in\mathbb{Z}$ and $\Delta\xi\in(-\pi,\pi]$ \cite{ruben2010vortex}. The value of $p$ corresponds to the detected charge numbers: +1, 0, and -1. +1 corresponds to a positive topological defect, while -1 corresponds to a negative topological defect. 

\subsection{Characterization of structural rearrangements}
The characterization of structural arrangements is performed using the well-established concept of $D_{\text{min}}^2$ \cite{falk1998nonaffine}, that allows for a direct identification of regions with large non-affine displacement.

In order to do that, we define 
\begin{equation}
\begin{aligned}
D^2(t,\Delta t)&=\sum_n\sum_i\left(r_n^i(t)-r_0^i(t)-\sum_j(\delta_{ij}+\varepsilon_{ij})\right.\\&\times[r_n^j(t-\Delta t)-r_0^j(t-\Delta t)]\biggr)^2,
\end{aligned}
\end{equation}
where the indices $i$ and $j$ represents spatial coordinates, while $n$ indexes the particles within the interaction range of the reference particle, with $n=0$ being the reference particle. $r_n^i(t)$ denotes the $i$th component of the position of the $n$th particle at time $t$. The $\varepsilon_{ij}$ that minimizes $D^2$ is then determined by calculating 
\begin{equation}
\begin{aligned}
X_{ij} = \sum_{n} \big[r_{n}^{i}(t)-r_{0}^{i}(t)\big]\times\big[r_{n}^{j}(t-\Delta t)-r_{0}^{j}(t-\Delta t)\big],
\end{aligned}
\end{equation}
\begin{equation}
\begin{aligned}
Y_{ij} & =\sum_n\left[r_n^i(t-\Delta t)-r_0^i(t-\Delta t)\right] \\
 & \times[r_n^j(t-\Delta t)-r_0^j(t-\Delta t)],
\end{aligned}
\end{equation}
\begin{equation}
\begin{aligned}
\boldsymbol{\varepsilon}_{ij}=\sum_kX_{ik}Y_{jk}^{-1}-\delta_{ij} .
\end{aligned}
\end{equation}
The minimum value of $D(t,\Delta t)$ defines $D_{\text{min}}^2$ that quantifies the local deviation from affine deformation during a time interval $[t-\Delta t,t]$. $D_{\text{min}}^2$ is an excellent diagnostic for identifying local irreversible deformations and structural rearrangements.

\clearpage
% Bibliography
%\bibliography{main}

% Produces the bibliography via BibTeX.


\clearpage

\section*{Supplementary Information}\label{Supplementary Information}

\color{red}
\subsection*{Measurement of restitution coefficient}
To measure the restitution coefficient, particle A was fixed sideways on the substrate, while particle B, also placed sideways, was released from a height $h_1 = 20\,\mathrm{cm}$ directly above A. After the head-on side-surface collision, the rebound height $h_2$ of particle B was recorded with a high-speed camera. Repeated trials with different particle pairs yielded the average rebound height, from which the restitution coefficient was obtained as  
\begin{equation}
\epsilon = \frac{|v_{2}|}{|v_{1}|} = \frac{\sqrt{2gh_{2}}}{\sqrt{2gh_{1}}} = \sqrt{\frac{h_{2}}{h_{1}}},
\end{equation}
where $v_1$ and $v_2$ are the pre- and post-collision velocities, and $g$ is the gravitational acceleration.

\subsection*{Single particle dynamics}

In our experiments, the persistence time $\tau_p$ is defined as the cutoff time marking the end of the initial superdiffusive regime. Specifically, $\tau_p$ corresponds to the time point at which the log-log mean squared displacements (MSD) slope first decreases below 1 after exceeding it:

\begin{equation}
k^* = \frac{d(\log(\mathrm{MSD}))}{d(\log(t))}, \qquad \tau_p = \max \left( t(k^* > 1) \right).
\end{equation}

\begin{figure*}[p]
   \centering
    \includegraphics[width=\linewidth]{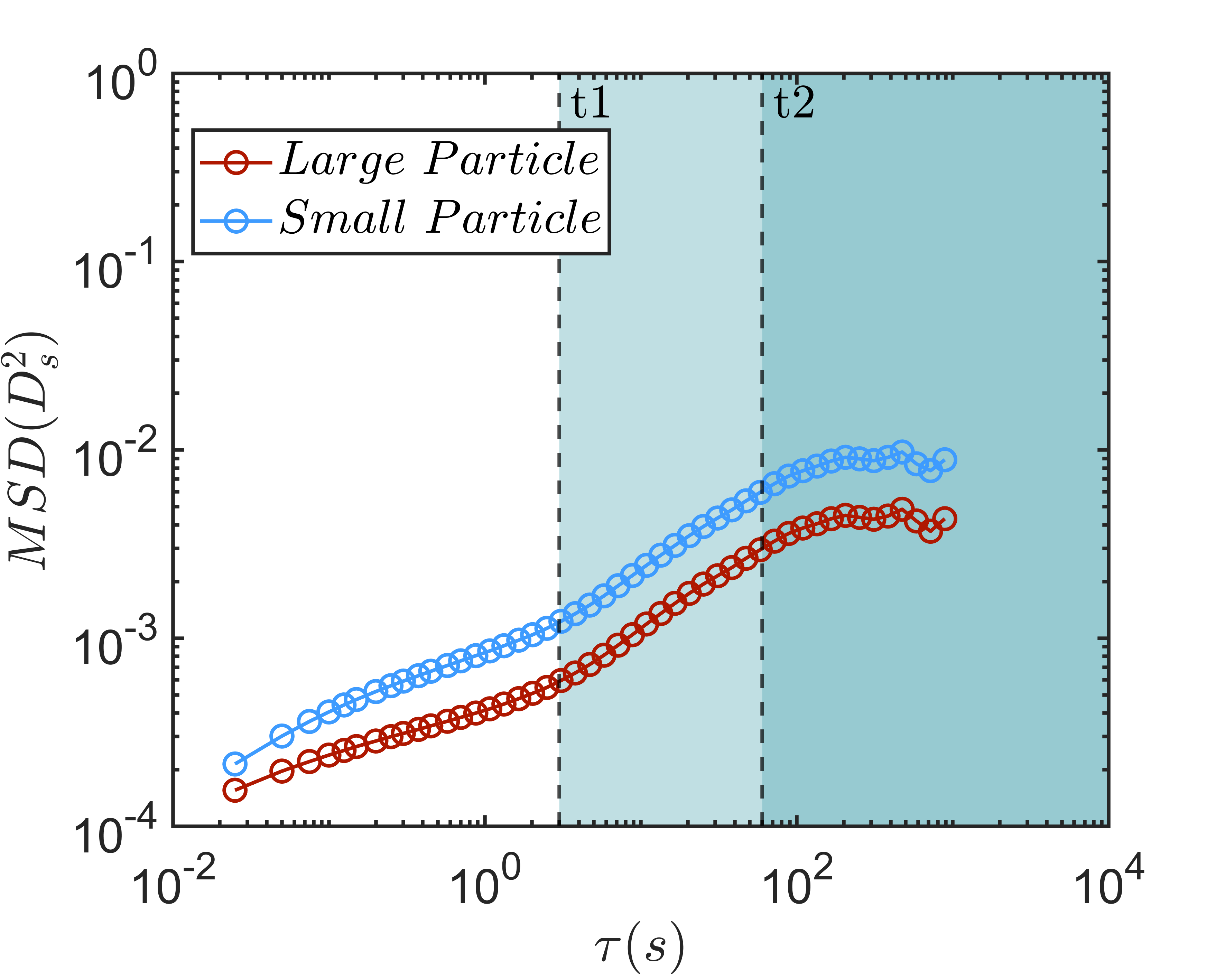}
    \caption{\label{figR5} Mean square displacements (MSD) for large and small particles at packing fraction $\phi=0.822$, in units of the square of the small particle
    diameter $D_s^2$. The background colors correspond to the three dynamical stages shown in Fig.~4 of the main text.}
\end{figure*}

It is worth noting that $\tau_p$ is closely related to the packing fraction 
$\phi$ due to interparticle interactions, as the frequency of interparticle collisions governs the persistence time. At the packing fraction of $\phi=0.822$, the MSD of large and small particles are presented in the Fig.~\ref{figR5}.

\begin{figure*}[htbp]
   \centering
    \includegraphics[width=\linewidth]{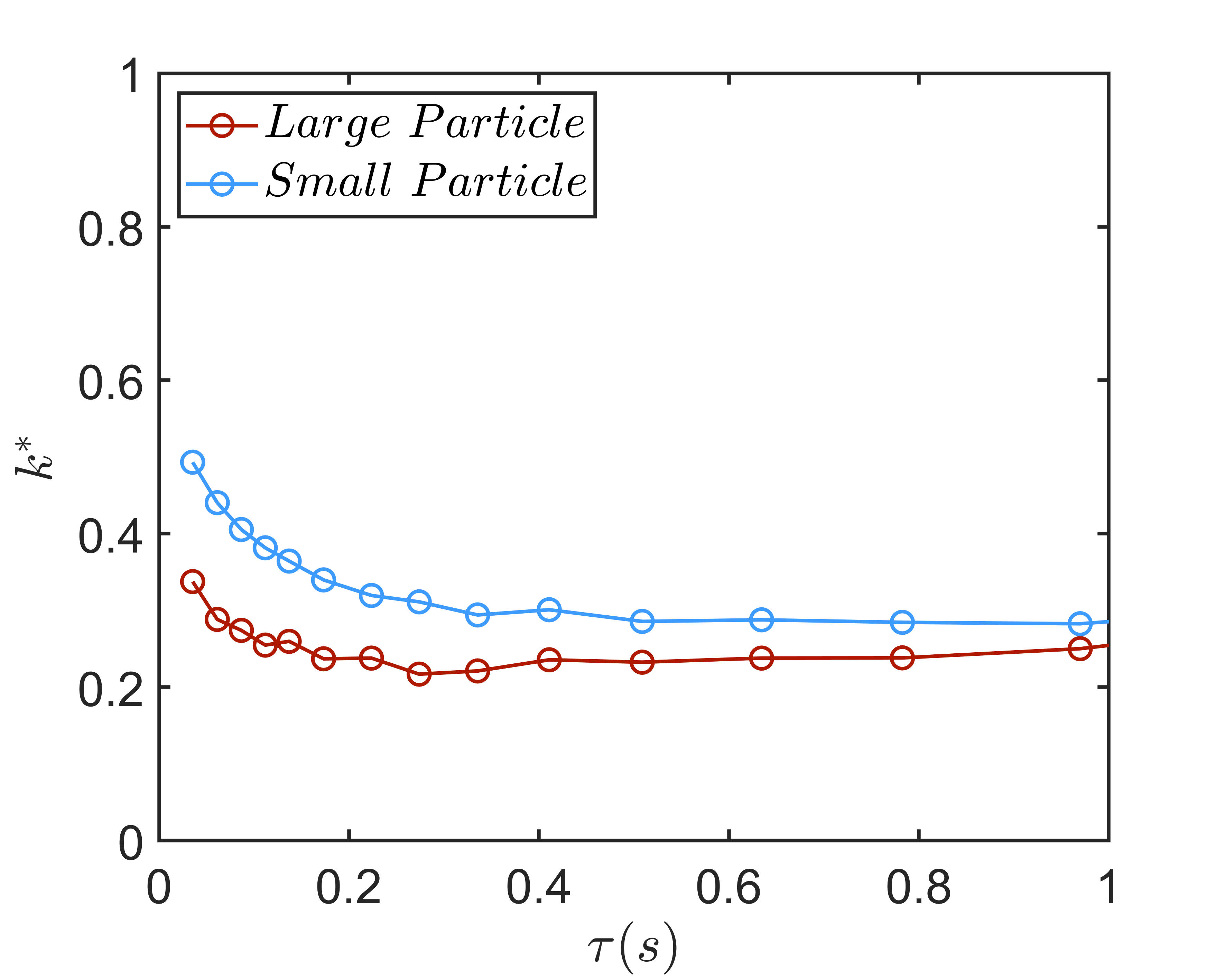}
    \caption{\label{figR6} The local slope $k^*$ of the mean square displacements (MSDs) curves in Fig.~\ref{figR5}.}
\end{figure*}

From this analysis, we can obtain the local slope of the MSD for large and small particles respectively in the log-log axis $k^*$, as shown in Fig.~\ref{figR6}.

At this high packing fraction, the particles do not exhibit superdiffusive behavior within the temporal resolution of our experiment ($\Delta t = 0.025\,\mathrm{s}$). We therefore approximate the persistence time of both large and small particles to correspond with the temporal resolution limit of $0.025\,\mathrm{s}$ under these conditions. Thus, we conclude that
\begin{equation}
\tau_p=0.025\,\mathrm{s},
\end{equation}
for both large and small particles.

In the following, we introduce the calculation of the persistence length and the Péclet number. The particles' diameter is $D_l=22.4\,\mathrm{mm}$ for large particles and $D_s=16\,\mathrm{mm}$ for small particles.
For both type of particles we estimate the root mean square active velocity $v_0$ from single-particle experiments:
\begin{equation}
v_{0,l} = \sqrt{\langle \overline{\mathbf{v}^2(t)} \rangle} \approx 1.58\,D_s \cdot \mathrm{s}^{-1} \approx 25.3\,\mathrm{mm} \cdot \mathrm{s}^{-1},
\end{equation}
\begin{equation}
v_{0,s} = \sqrt{\langle \overline{\mathbf{v}^2(t)} \rangle} \approx 1.43\,D_s \cdot \mathrm{s}^{-1} \approx 22.9\,\mathrm{mm} \cdot \mathrm{s}^{-1}\,\,.
\end{equation}
Here, the subscripts $l$ and $s$ respectively represent large and small particle. The notation $\langle \cdots \rangle$ denotes an average over repeated experimental realizations, while the overline indicates a time average along individual particle trajectories.

Then, the persistence length $l_p$ is 
\begin{equation}
    l_{p,l} = v_{0,l} \cdot \tau_p = 25.3\,\mathrm{mm/s} \times 0.025\,\mathrm{s} = 0.6325\,\mathrm{mm},
\end{equation}
\begin{equation}
    l_{p,s} = v_{0,s} \cdot \tau_p = 22.9\,\mathrm{mm/s} \times 0.025\,\mathrm{s} = 0.5725\,\mathrm{mm}\,.
\end{equation}

Finally, the Péclet number $\mathrm{Pe}$ is
\begin{equation}
\mathrm{Pe}_l = \frac{l_{p,l}}{D_l} = 25.3\,\mathrm{mm/s} \times 0.025\,\mathrm{s} \times \frac{1}{22.4\,\mathrm{mm}} = 0.0282,
\end{equation}
\begin{equation}
\mathrm{Pe}_s = \frac{l_{p,\mathrm{s}}}{D_\mathrm{s}} = 22.9\,\mathrm{mm/s} \times 0.025\,\mathrm{s} \times \frac{1}{16\,\mathrm{mm}} = 0.0358,
\end{equation}
where the subscripts $l$ and $s$ represent large and small particle respectively.

\subsection*{Discussion of the effect of system boundaries on the final vortex }

The flower-shaped boundary is expected to have a negligible impact on the observed vortex field. This is supported by the observation that, when large internal vortices form, the displacements of boundary particles remain small and exhibit disordered directions, indicating minimal coupling with the internal vortex motion.

%The boundary was intentionally designed to suppress collective rotation along the edges and to facilitate reinjection of particles near the boundary back into the system.

For comparison, in a monodisperse system at high packing fraction, where crystallization occurs, the boundary geometry can influence flocking motion due to long-range translational order. In contrast, in our bidisperse, structurally disordered system, such boundary effects are expected to be negligible.
\color{black}

\subsection*{Spatial structure of topological defects}
We characterize the spatial structure of the topological defects by investigating their radial pair correlation functions $g_{\alpha \beta}(r)$, where $\alpha,\beta=\pm$ indicate the sign of the corresponding winding number.

The radial pair correlation function between topological defects with $\alpha$ and $\beta$ charges is defined as
\begin{equation}
    g_{\alpha \beta}(r) = \frac{L_xL_y}{2\pi r N_\alpha N_\beta} \sum_{i=1}^{N_\alpha} \sum_{j=1}^{N_\beta} \delta(r-|\Vec{r_{ij}}|),
\end{equation}
 where $L_x, L_y$ are the dimensions of the region in which correlation is calculated, $N_{\alpha \beta}$ is the number of defects with $\alpha,\beta$ charge inside the region, and $|\Vec{r_{ij}}|$ is the distance between topological defects. 
 
\begin{figure*}[htbp]
\centering
\includegraphics[width=\linewidth]{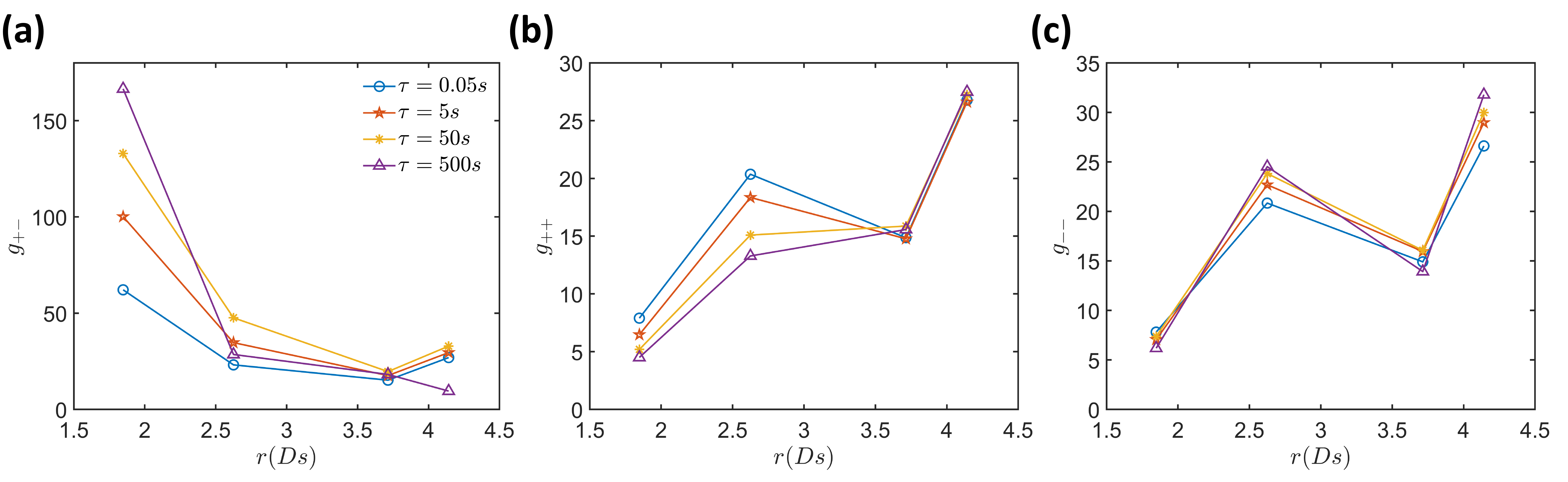}
\caption{\label{S1} \textbf{Spatial correlation of topological defects. } Pair correlation function of positive and negative defects \textbf{(a)}, positive and positive defects \textbf{(b)}, negative and negative defects \textbf{(c)} for different values of the observation time $\tau$.}
\end{figure*}

We display our results in Fig. \ref{S1}. First, we observe that at any observation time, TDs with opposite charge shows a strong short-range correlation, with a large value of $g_{\pm}$ at the first lattice point. This indicates that opposite-sign defects tend to stay close to and attract each other, forming pairs with zero total winding number. On the other hand, the correlations between TDs with same charge do not show a peak at short length-scales.  In contrast, as shown in Fig. \ref{S1}(b) and (c), for defects of the same sign, the value at the first lattice point is low, suggesting that same-sign defects tend to repel each other. 

Moreover, as shown in Fig. \ref{S1}(a), the short range correlation between $+$ and $-$ TDs becomes more pronounced at late time, with the first peak in $g_{\pm}$ being much larger for long $\tau$. This suggests that, while the system relaxes towards the final non-equilibrium steady state, the TDs self-organize in short-range pairs and display a more ordered spatial distribution.

%On the other hand, the deviation in $g_{+-,-+}$ are not so significant upon moving $\tau$.

\subsection*{Relaxation dynamics and dynamical heterogeinities}

In order to further study the dynamics of our system, we consider two typical probes: the intermediate scattering function $F_s(q,t)$ and the four-point susceptibility $\chi_4(t)$.

\begin{figure*}[htbp]
\centering
\includegraphics[width=\linewidth]{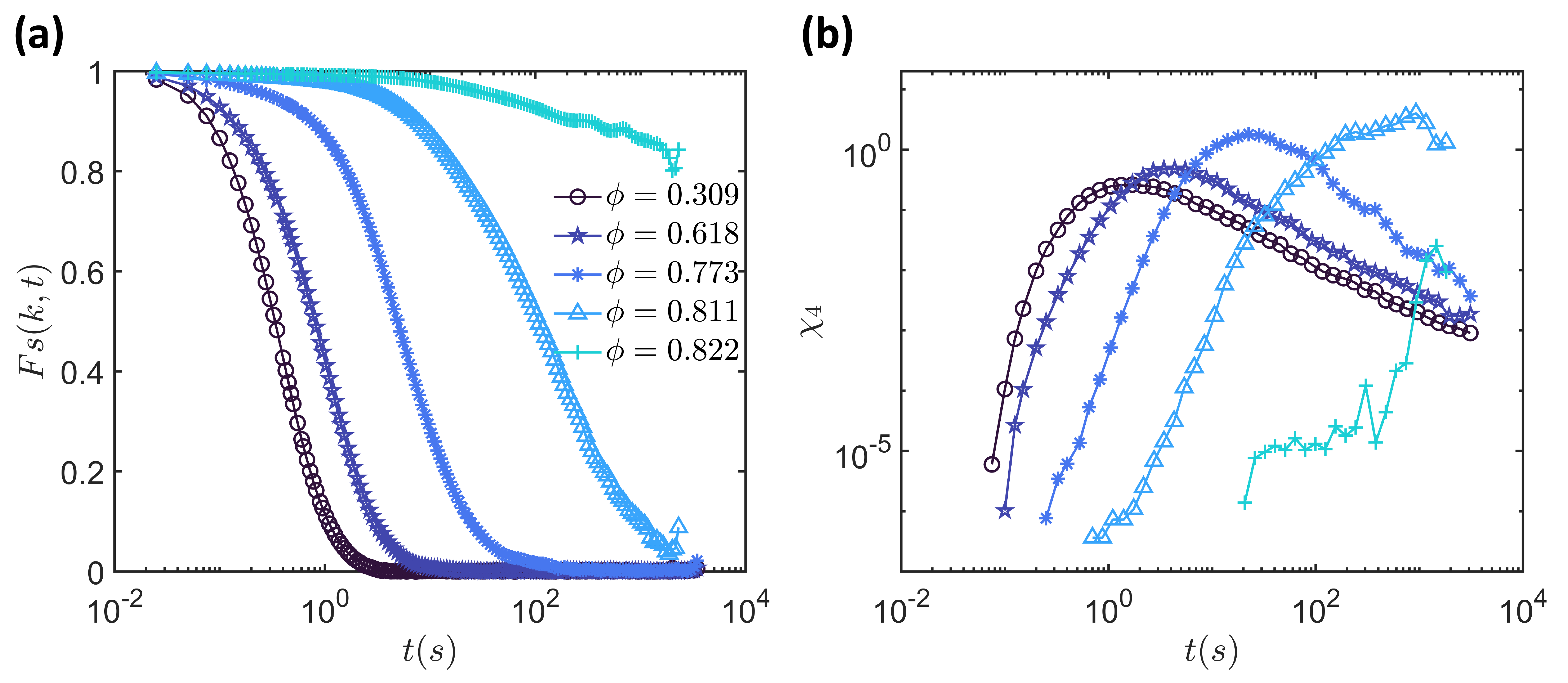}
\caption{\label{S2} \textbf{Dynamical characterization. } Intermediate scattering function $F_s(q,t)$ as a function of time for different packing fractions $\phi$ \textbf{(a)}. Four point susceptibility function $\chi_4(t)$ as a function of time for the same values of the packing fraction $\phi$ \textbf{(b)}. Same color scheme is adopted in both panels of this figure.}
\end{figure*}

\begin{figure*}[htbp]
   \centering
    \includegraphics[width=\textwidth]{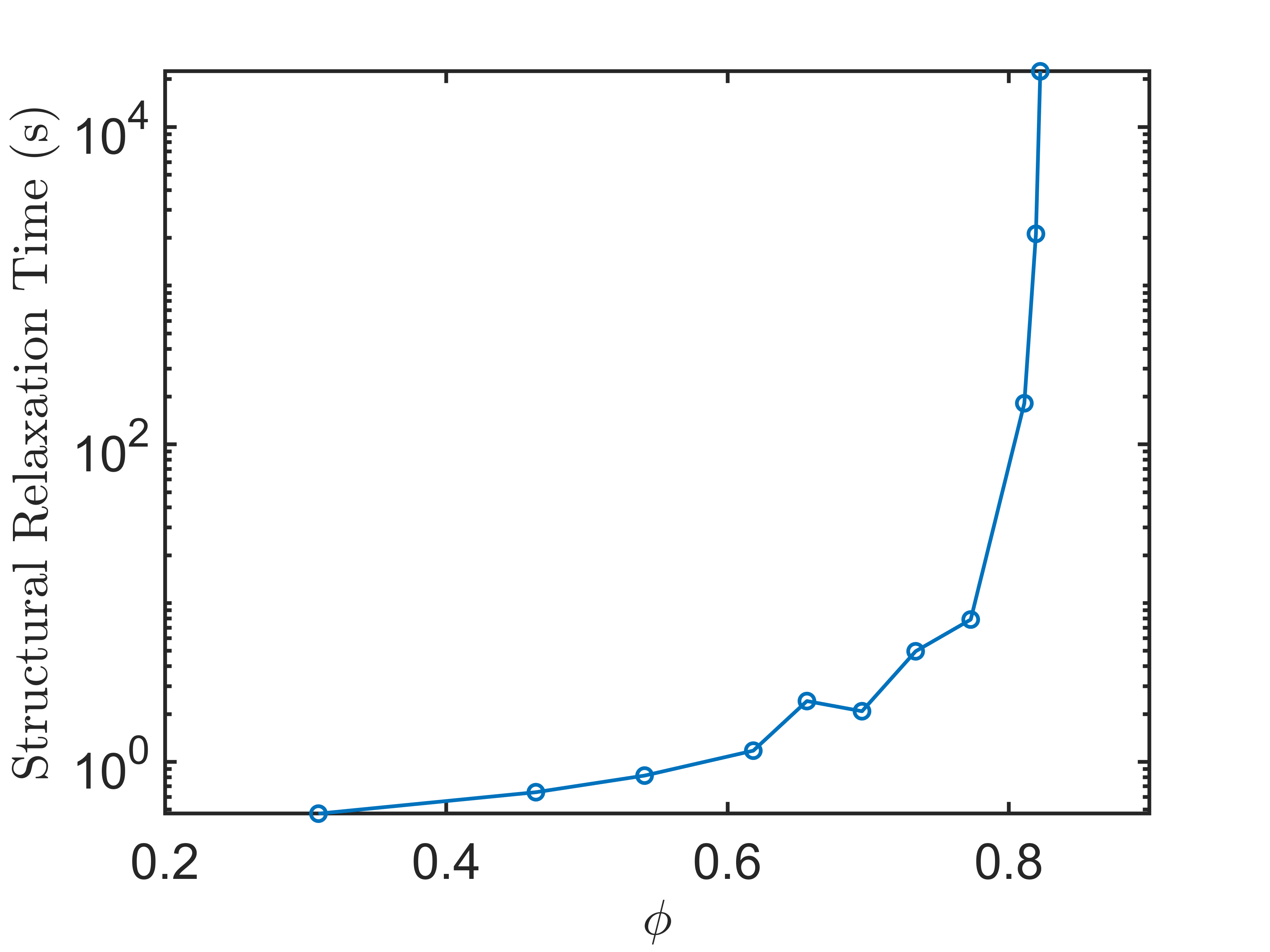}
    \caption{\label{figR7} Structural relaxation time versus packing fraction $\phi$. The structural relaxation time is determined from the self-intermediate scattering function as the time at which it decays to $1/e$.}
\end{figure*}

The intermediate scattering function is defined as
\begin{equation}
    F_s(q,t)=\frac{1}{N}\Sigma_{j=1}\left\langle e^{-i\vec{q}\cdot(\vec{r}_j(t)-\vec{r}_j(0))}\right\rangle,
\end{equation} 
where $q$ is the wave vector, $N$ is the total number of particles in the system, $r_j(t)$ is the position of particle $j$ at time $t$. Results in Fig. \ref{S2}(a) show that system exhibits progressively slower relaxation by increasing packing fraction. {\color{red} The structural relaxation time $\tau_a$ as a function of packing fraction $\phi$ is presented in Fig.~\ref{figR7}. Here, $\tau_a$ is defined via the self-intermediate scattering function as the time at which it decays to $1/e$.}

The four-point susceptibility in  Fig. \ref{S2}(b) describes the degree of dynamical heterogeneity in space within a time interval, and is defined as: $\chi_4(t)=N[\langle q_s(t)^2\rangle-\langle q_s(t)\rangle^2]$ , where $q_s(t)=(1/N)\sum_{i=1}^{N}w(|r_{\mathrm{i}}(t)-r_{\mathrm{i}}(0)|)$ , $w=1(0)$ if $|r_{\mathrm{i}}(t)-r_{\mathrm{i}}(0)|<(>)0.5D$. $N$ is the total number of particles in the system, $r_i(t)$ is the position of particle $i$ at time $t$, and $D$ is the average particle diameter. The average $\langle \cdot \rangle$ is performed over all particles and all initial times. 
The peak position in $\chi_4(t)$ indicates the time during which the dynamics exhibit the greatest heterogeneity, while the peak height reflects the intensity of these heterogeneities. As the packing fraction increases, the time at which the system reaches its maximum dynamical heterogeneity is delayed. Meanwhile, from a packing fraction of $\phi=0.309$ to $0.811$, the peak intensity of dynamical heterogeneity also increases. Due to experimental limitations, the data for $\phi=0.822$ do not exhibit a distinct peak within the accessible observation time. However, we speculate that a peak would emerge over a longer time scale. 

\color{red}
\subsection*{Characterization of three stage non-equilibrium dynamics}

\begin{figure*}[htbp]
   \centering
    \includegraphics[width=\linewidth]{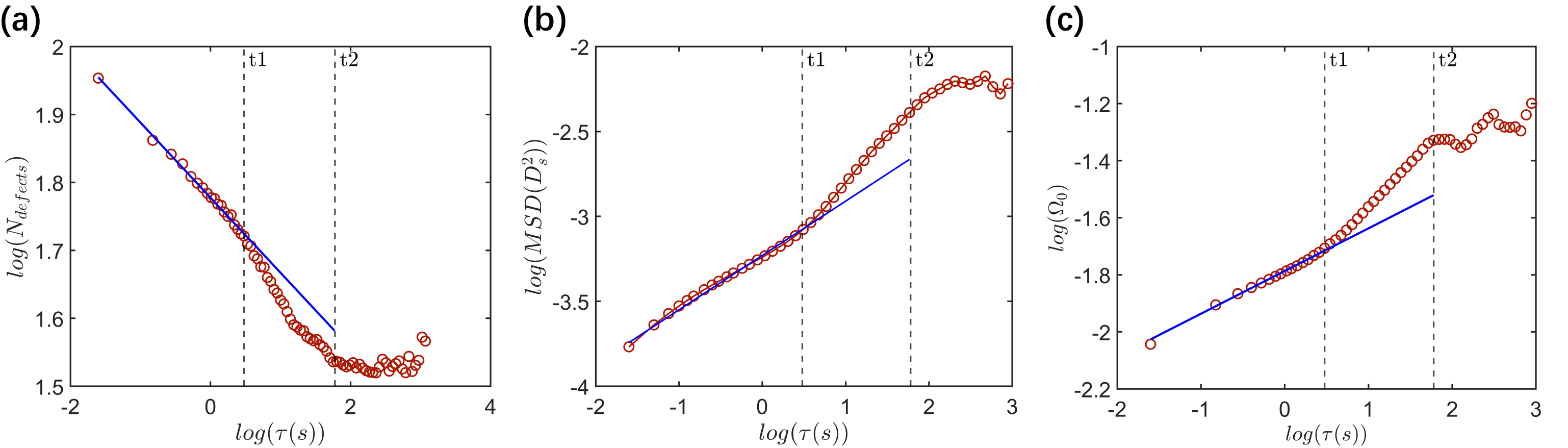}
    \caption{\label{figR4} Characterization of the three-stage non-equilibrium dynamics. \textbf{(a)} Topological signatures. \textbf{(b)} Particle-level dynamics. (\textbf{c)} Collective motion. In all panels the solid blue line is a power-law fit of the experimental data in the range $t<t_1$ extended up to $t=t_2$. The deviation of the experimental data from this power-law behavior at $t_1$ is evident in all three physical quantities.}
\end{figure*}

To highlight the transition at $t = t_1$, we fitted the data for $t < t_1$ with a power-law function. As shown in Fig.~\ref{figR4}, while the data follow the fit closely for $t < t_1$, clear deviations arise for $t > t_1$, indicating a distinct change in the dynamical behavior.
\color{black}

%%% Each figure should be on its own page

%%% Add this line AFTER all your figures and tables

\end{document}